\documentclass[]{aa}
\usepackage[latin1]{inputenc}
\usepackage{graphicx}
\usepackage{txfonts}
\usepackage{natbib}
\bibpunct{(}{)}{;}{a}{}{,}

\newcommand{\dpart}[2]{\frac{\partial #1}{\partial #2}}

\begin{document}

\title{Effects of turbulent diffusion  on the chemistry of
diffuse clouds}

\author{P. Lesaffre \inst{1}
\and
M. Gerin \inst{1}
\and
P. Hennebelle \inst{1}
}

\institute{LERMA, CNRS UMR8112, Observatoire de Paris and ENS/LRA, 24 Rue Lhomond, 75231 Paris cedex 05, France}

\date{Received September 15, 1996; accepted March 16, 1997}

\abstract 
{} 
  {We probe the effect of turbulent diffusion on the chemistry at the
interface between a cold neutral medium (CNM) cloudlet and the warm
neutral medium (WNM).}
  {We perform moving grid, multifluid, 1D, hydrodynamical simulations
with chemistry including thermal and chemical diffusion. The diffusion
coefficients are enhanced to account for turbulent diffusion. We
post-process the steady-states of our simulations with a crude model
of radiative transfer to compute line profiles.  }
  {Turbulent diffusion spreads out the transition region between the
CNM and the WNM. We find that the CNM slightly expands and heats up:
its CH and H$_2$ content decreases due to the lower density.  The
change of physical conditions and diffusive transport increase the
H$^+$ content in the CNM which results in increased OH and
H$_2$O. Diffusion transports some CO out of the CNM. It also brings
H$_2$ into contact with the warm gas with enhanced production of
CH$^+$, H$_3^+$, OH and H$_2$O at the interface.  O lines are
sensitive to the spread of the thermal profile in the intermediate
region between the CNM and the WNM.  Enhanced molecular content at the
interface of the cloud broadens the molecular line profiles and helps
exciting transitions of intermediate energy. The relative molecular yields
are found higher for bigger clouds.}
   {Turbulent diffusion can be the source of additional molecular
production and should be included in chemical models of the
interstellar medium (ISM). It also is a good candidate for the
interpretation of observational problems such as warm H$_2$, CH$^+$
formation and presence of H$_3^+$.}


\maketitle

\section{Introduction}

  In the thermally bistable neutral ISM, small structures of cold
neutral gas (CNM for cold neutral medium, T $\approx 70$~K and $n
\approx $50~cm$^{-3}$ ) are immersed in the warm neutral medium (WNM, T
$\approx 8000$K and $n \approx 0.4$ cm$^{-3}$).  Both phases are found
to have approximately the same pressure \citep{HT03}. In this paper, we
investigate how turbulent diffusion processes can affect the chemistry
and the structure of an interface between a CNM cloudlet and the WNM.

  Much of the knowledge on the cold diffuse medium has been
established from the detection of molecules, radicals and molecular
ions.  Despite its relatively low density the diffuse interstellar
medium is chemically active and furthermore, its chemistry is not
fully understood \citep[][ and references therein]{SM06}.  The
presence of the reactive ions CH$^+$ and H$_3^+$ together with
radicals like CH and C$_2$ and molecules like CO in the diffuse ISM
can not be understood with steady state chemical models. For example,
\cite{LP04} present a detailed modeling of the ISM along the line of
sight of $\zeta$ Per  which requires three components along the line
of sight : a long (4 pc) diffuse region at 60 K, a tiny (100 AU) dense
region at 20 K, both with a large ionization rate, and a series of
shocks which account for the abundance of CH$^+$ and the excited
rotational populations of H$_2$.  This example is not unique and
demonstrates the necessity of coupling self-consistent models of the
structure of the ISM with the chemistry.  It also shows that CH$^+$
and H$_3^+$ are key species for understanding the coupling of
turbulence, dynamics and chemistry in the diffuse ISM.

The ubiquitous presence of CH$^+$ and excited rotational lines of
H$_2$, which are not expected at thermal equilibrium given the average
temperatures of the cold neutral medium, indicates that non
equilibrium processes must also be taken into account, which will
bring enough energy locally to trigger the formation of CH$^+$ and
could contribute to the excitation of H$_2$. H$_3^+$ is mainly formed
with H$_2$ and H$_2^+$ (itself resulting from the ionisation of H$_2$
by a cosmic ray) and efficiently destroyed by dissociative
recombination with electrons. As reviewed by \cite{SM06}, diffuse
cloud models do not predict H$_3^+$ to be abundant, except when the
ionisation rate is high enough to maintain a large formation
rate. Steady state chemistry predicts that the H$_3^+$ density scales
as: $n($H$_3^+)=$$ {\zeta n(H_2)} \over {k_e n(e)}$. Because the
H$_3^+$ density is approximately constant and proportional to $\zeta$,
the presence of large H$_3^+$ column densities requires long
pathlengths through the CNM (4 pc for $\zeta$~Per as modelled by Le
Petit et al. 2004 and using $\zeta = 25 \times 10^{-17} \ s^{-1}$)
which significantly exceed the expected size of such structures
\citep[e.g. around 0.1~pc up to 1~pc in the simulations by][]{AH05}.
The problem with the path length is even more severe because the
determined column density of H$_3^+$ towards $\zeta$ Per requires a
path length larger than 60~pc when using the standard value for the
cosmic ray ionisation rate , $\zeta \sim 3 \times 10^{-17} \ s^{-1}$ !
As CNM cloudlets are never isolated but always immersed in the WNM, we
investigate in this paper the effect of turbulent diffusion at their
interface as an alternative solution for enhancing molecular
abundances in the ISM.  A subsequent paper will present the effects of
shocks.

  The chemical importance of turbulent diffusion has been acknowledged
in the context of protostellar cores since the work of \cite{BD82} who
realised how extra mixing could limit element depletion on
grains. Since then, many authors have computed the effect of turbulent
mixing on the chemistry of protostellar and prestellar
cores. \cite{CP89} and \cite{C91} have used a simple two-zone model
and a mixing prescription between the envelope and the cores of these
objects.  \cite{X94} and \cite{X95} computed the isothermal evolution
of chemistry with an enhanced diffusion coefficient. \cite{W02} later
extended this work including H$_2$ photo-dissociation and gas-grain
interactions. \cite{RH97} did similar computations but using a uniform
pressure and a diffusion term based on numerical densities rather than
a Fick law based on mass fraction.

  In the present study, we transpose this series of work to another
context with physical conditions at lower densities and lower visual
extinctions, more appropriate to the diffuse neutral ISM.  We also
improve on the previously used numerical methods.  First, we compute
the hydrodynamical evolution of a multifluid gas and we treat the gas
cooling in a self-consistent manner, completely coupled to the
chemical evolution.  Second, we model the turbulent diffusion of
temperature along with chemical mixing: this will prove useful to
extend the region where the formation of molecules is favoured.

  In section 2 we describe our numerical method. We present our
results in section 3 and discuss them in section 4. Section 5 draws
our conclusions.

\section{Method}

We use the multifluid monodimensional (plane parallel)
magnetohydrodynamical (MHD) code presented in \cite{L04a}. This code
fully couples the hydrodynamics and the chemistry. It also
makes use of a moving grid algorithm \citep{DD87} which allows to
spatially resolve the shock discontinuities and the diffusion fronts.
In order to model the diffuse molecular gas we had to supplement the
code with the following new processes.

\subsection{Radiative transfer}

  A radiation field equal to the standard interstellar radiation field
(ISRF) illuminates the {\it right hand} (outer) side of the
computational box.  The extinction at the outer side of the box is
very small: $A_{v0}=0.01$, relevant to WNM conditions.  The
extinction at any point inside the computational box is given by

\begin{equation}
A_v(r)=A_{v0}+1.5 \times 10^{-21}{\rm cm}^2 \int_r^R N_H {\rm d}l
\label{Av}
\end{equation}
where $r$ is the position inside the computational box, $R$ is the
position at the outer (right hand) edge of the box and $N_H$ is
the numerical density of hydrogen nuclei (hence $A_v$ actually
increases from right to left).  In most simulations of this paper,
the total $A_v(0)$ across the box is $0.14$ mag for a total length of
the box $R=5$~pc: we refer to them as ``Small cloud'' (or standard)
simulations. In some other simulations, we used $A_v(0)=1.4$ mag and
$R=10$~pc (``Big cloud'' simulations). The local value of $A_v$ is
used to determine the heating rate due to the photo-electric effect on
grains and various photo-chemical rates (see below).

 The dense cloud will therefore naturally be formed at the left
hand side of the computational box, where the extinction is the
highest. The WNM will be found at the right hand side.

\subsection{Photo-chemistry}

  The reaction network in \cite{L04a} has been extended with
photo-chemical reactions of dissociation and ionisation. The total
number of reactions is now 138. The full list of reactions can be
found in the last appendix of \cite{L02} and takes its main roots in
the early work of \cite{CPF98}: it is designed to account for the
abundances of the main cooling agents of the ISM such as C$^+$, C, O,
H$_2$, OH, CO and H$_2$O.  The dependence of each photo-reaction rate
on the local radiation field is modeled with a parameter $\beta$. The
local rate is obtained by multiplying the rate for the ISRF by the
factor $\exp(-\beta A_v)$. In most cases, this simple rule is
sufficient to get an accurate model for the extinction of the
radiation field in the wavelength range relevant to a given
photo-reaction.

  However, the photo-dissociation of the molecule H$_2$ requires a
special treatment because of its self-shielding properties. For this
purpose, we use equation (37) of \cite{DB96} with a constant Doppler
parameter of $b_5=0.1$ corresponding to a 0.1~km.s$^{-1}$ Doppler
broadening (this small value maximises the effect of
self-shielding). Note that due to the small total column-density
across our computational box, we do not need a similar treatment for
the photo-dissociation of CO.

  In the present simulations we use $1.5~10^{-7}$~cm$^{-3}$s$^{-1}$ at
300~K for the dissociative recombination rate of H$_3^+$ with a cosmic
ray ionisation rate of $\zeta=5~10^{-17}$~s$^{-1}$.

\subsection{Diffusion}
  The structure and evolution of a front between a CNM cloudlet
surrounded by the WNM are mainly determined by diffusion
processes. Diffusion is due to random motions which transport thermal energy
or chemical species. It can be caused either by microscopic collisions
between molecules or by random macroscopic motions of turbulent
eddies.

  In the following, we evaluate diffusion coefficients
for these two modes of diffusion.

\subsubsection{Molecular diffusion}
\label{turb}
 It is a difficult task, still heavily discussed, to determine the
diffusion coefficients in a multi-component plasma. Here, we adopt a
simple view consistent with our modeling of three different
fluids: neutrals, ions and electrons. 

A typical molecular flight takes place at the thermal velocity over a
length scale equal to the mean free path. We hence adopt diffusion
coefficients of the form
\begin{equation}
 d_j=c_j \lambda
\end{equation}
with $j=$n,i,e spanning the neutral, ion and electron fluids, where
 $\lambda=1/\sigma N$ is the mean free path ($\sigma=10^{-15}$~cm$^2$
 and $N$ is the total number density of particles per cm$^3$) and
\begin{equation}
c_j=\sqrt{\frac{5 {\rm k} T_j}{3 \mu_j}}
\end{equation}
is the sound speed in fluid $j$ (k is the Boltzmann constant, $T_j$ is
the temperature of fluid $j$ and $\mu_j$ its average molecular
weight). We assume that electrons are dynamically coupled to the ions
and hence take $d_e:=d_i$. According to the relative average molecular
weights, diffusion is fastest in the neutral fluid and slowest in
the ion and electron fluid. It is a bit artificial to adopt the
same mean-free path for all three components independent of energy,
but as long as these length scales are much smaller than the turbulent 
diffusion lengths scales, our results should not be changed.

The thermal diffusion term takes the form 
\begin{equation}
\dpart{}{r}(\widetilde{N_jd_j}\dpart{{\rm k} T_j}{r})
\end{equation}
in the evolution equation for the energy density for each fluid
$j$. $N_j$ is the number density of particles in fluid
$j$. $\widetilde{N_jd_j}$ is $N_jd_j$ interpolated linearly at each
interface of the computational zones and diffusion fluxes are assumed
to vanish at the computational boundaries.

We adopt a similar treatment for the chemical diffusion term
\begin{equation}
\dpart{}{r}(\widetilde{\rho d_j}\dpart{X_i}{r})
\end{equation}
in the evolution equation for the number density $N_i$ of species $i$
in fluid $j$, with $X_i$ the mass fraction of species $i$ and $\rho$
the total mass density. Note that we do not take into account the
individual thermal velocities for each species but we rather take the
average sound speed in each corresponding fluid. This may be improved
in future work.

\subsubsection{Turbulent diffusion}
Consider the evolution equation of the number density $N$ of a given
chemical species (of mass fraction $X$) with velocity ${\bf U}$,
net chemical production rate $R$ and diffusion coefficient $d$:
\begin{equation}
\label{cheminf}
\dpart{N}{t}+{\bf \nabla.}(N{\bf U})=R+ 
{\bf \nabla.}[d\rho{\bf \nabla} X] \mbox{.}
\end{equation}
  Our purpose is to find an evolution equation for the quantity
$N$ averaged over a given length scale $l$, which we denote $\overline{N}$.
  The average of equation (\ref{cheminf}) yields
\begin{equation}
\label{chemav}
\dpart{\overline{N}}{t}+{\bf \nabla.}(\overline{N}~\overline{{\bf U}})=
\overline{R}+ {\bf \nabla.}[\overline{d\rho{\bf \nabla} X}]
-{\bf \nabla.}\overline{(n {\bf u})}
\end{equation}
 where ${\bf u}={\bf U}-\overline{{\bf U}}$ and $n=N-\overline{N}$.

 The challenge of turbulent models resides in the difficulty of
finding a simple but accurate model for the behaviour of high order
moments such as ${\bf \nabla.}\overline{(n {\bf u})}$. We now sketch a brief
justification for the crude model we adopted. We compute the
time evolution of $n$ under the first order smoothing approximation (FOSA)
which allows us to drop all orders higher than two. The difference
between equations (\ref{cheminf}) and (\ref{chemav}) yields
\begin{equation}
\label{equnt}
\dpart{n}{t}+{\bf \nabla.}(n{\bf U})=
-{\bf \nabla.}(\overline{N}{\bf u})+(R-\overline{R}) 
\end{equation}
where we neglected second order terms and safely dropped the term
proportional to $d$ since it is generally small compared to the
turbulent diffusion. The term $R-\overline{R}$ arises because of
differential reactivity in turbulent eddies \citep[see][]{L05} and can
be neglected provided the turbulent mixing time scales are much
shorter than the chemical time scales. That is hard to justify, and
can in fact be wrong for most of the rapidly reacting species, but we
will neglect this term in the present work for the sake of
simplicity. Note that we also make the convenient approximation
$\overline{R(N)}=R(\overline{N})$ which is consistent with our neglect of
differential reactivity.

  We now assume that $n$ is given by the integration of the source
term in equation (\ref{equnt}) over a given correlation time $\tau$,
sufficiently short to neglect the time evolution of spatially averaged
values.  This leads us to the expression for the turbulent term in
equation (\ref{chemav}):
\begin{equation}
-{\bf \nabla.}\overline{(n{\bf u})}={\bf \nabla.}[\overline{
\tau{\bf \nabla.}(\overline{N}{\bf u}){\bf u}}] \mbox{.}
\end{equation}

  Mass conservation of the averaged quantities requires
  $\overline{N}=\overline{\rho}~\overline{X}$. We use this relation to
  expand the previous relation:

\begin{equation}
-{\bf \nabla.}\overline{(n{\bf u})}=
{\bf \nabla.}[\overline{\rho}~\overline{(\tau {\bf u \otimes u})}
{\bf .\nabla} \overline{X}]
+{\bf \nabla.}[\overline{X}~\overline{(\tau {\bf u \otimes \nabla}){\bf .}
(\overline{\rho}~{\bf u})}]
\end{equation}

  We retain only the first term which is analogous to a diffusion term
with a tensor turbulent diffusion coefficient $K=\overline{ \tau {\bf
u \otimes u}}$ which simplifies to a scalar $K=\frac13\overline{\tau
u^2}$ for isotropic turbulence. The second term is an advection term
that is present only when turbulence carries a non zero net mass flux.

  Under these simplifying assumptions, the effect of an underlying
turbulent velocity field can hence be reduced to an effective
diffusion coefficient $K$. A good discussion of the observational
uncertainties on $K$ can be found in \cite{X95}. In brief, we have a
good idea of the rms fluctuations of the velocity field from the
broadening of spectral lines, but we only have a poor understanding of
the correlation length scale (or equivalently of the correlation time
scale) of the turbulence.

  Given these uncertainties, we investigated several possibilities for
the value of $K$.  For all computations with turbulent diffusion
presented in this paper, we keep the same formulation as for the
molecular diffusion coefficients $d_j$ above, but we use a fixed
correlation length $L$ instead of $\lambda$ and the sound speed $c_n$
in the neutrals for all fluids. This implicitly assumes that the
turbulent diffusion coefficient scales as the local average sound
speed, which may be questionable. On the other hand, it yields smaller
diffusion coefficients in colder parts of the cloud and it implies
that the ratio of the thermal pressure to the turbulent pressure
is a fixed constant, which seems reasonable (in particular, it gives
a uniform turbulent pressure field when the thermal pressure itself is
uniform). We also tested the more simple uniform prescription
$K=L\times 0.2$~km.s$^{-1}$ \citep[similar to ][]{X95} but we did not
find any significant qualitative difference with the results presented
here. In all our tests, we adopted the same diffusion coefficient for
the chemistry {\it and} the temperature.

\section{Results}

\subsection{Numerical setup}
  In this section, we investigate the steady state of condensation
fronts by means of time-dependent simulations evolved over very long
times.  We prepare a cold state (CNM) and a hot state (WNM) of
thermochemical equilibrium at a given total pressure $P_0=3\times
10^{3}$~K.cm$^{-3}$ for a given extinction $A_{v0}=0.01$. More
precisely, we evolve a two-zones isobaric model until both zones
achieve a steady state of thermal and chemical equilibrium at
$A_{v0}=0.01$, hence the initial molecular content is small. Our
adopted values for the abundance of C and O gas phase are
1.3~10$^{-4}$ and 4.3~10$^{-4}$ respectively.  We then select a length
$l$ ($l=0.5$~pc in for our small cloud simulations and $l=5$~pc for
our big cloud simulations) for the cold ``cloud'' which will be
located to the left (inner) side of the computational box and the rest
of the box (of total length $R$=5~pc or 10~pc in the small or big
cloud simulations) is filled with the hot state. We then let the gas
evolve chemically and hydrodynamically in the box.  This procedure
essentially sets only the total mass in the box, as the size of the
cloud and the pressure will later relax according to the
hydrodynamical, thermal and chemical evolution. We later refer to the
small cloud simulation without turbulent diffusion as our reference
run.

\subsection{Evolution of the small cloud without turbulent diffusion (reference run) }

The chemistry first reacts to the new local $A_v>A_{v0}$, with
molecules reforming in the cloud. Another transient effect is due to
the partial pressure of ions and electrons which differ in the cold
and hot phase due to fractional abundance differences.  The gradient
of partial pressures generates a flow of ions across the surface of
the cloud.  This process is responsible for a transient chemistry which we do
not describe here, as it mainly results from our arbitrary initial
conditions. Microscopic (molecular) thermal and chemical diffusion
then build up a smooth spatial transition between the cold and hot
phases.

\begin{figure}[h]
\centerline{
\includegraphics[width=7cm,angle=-90]{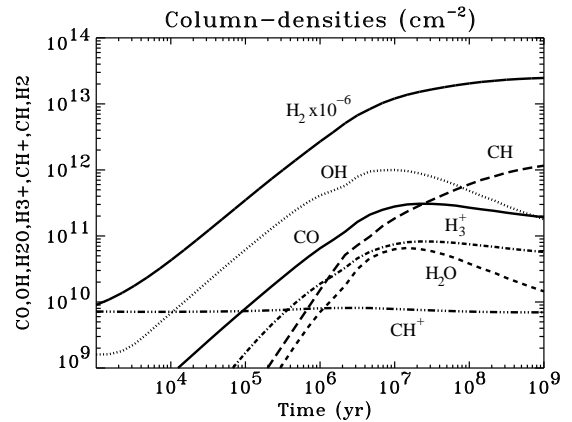}
} 
 \caption{Reference run (small cloud without turbulent diffusion):
time evolution of H$_2$ (solid, value scaled by a factor 10$^{-6}$),
CO (solid), OH (dotted), H$_2$O (dashed), H$_3^+$ (dash-dotted),
CH$^+$ (dash-dot-dot-dot) and CH (long-dash) column-densities
(cm$^{-2}$) integrated over the computational box.  }
\label{Nt}
\end{figure}

  We give the time evolution of the total column densities of several
species in figure \ref{Nt} as an illustration of the very long times
required to reach a steady state. One after the other, chemical
reactions reach their steady state according to their own time
scales. After a time of about $0.1$~Gyr chemical equilibrium is
achieved and the shape of the diffusion front (figure \ref{Tn}) reaches
an approximate steady-state. However, it is not stationary yet as it
is moving from the hot towards the cold medium, with the cold medium
reaching colder temperatures and higher molecular fractions. The cloud
shrinks in size as the condensation front moves and its density gets
higher while the hot medium gets hotter and more diffuse. Note that
the overall pressure rises while the front tries to adjust the total
cooling rate to the total heating rate, on its way to find a steady
state.

This simulation did not show any sign of reaching a steady state after
more than 1 Gyr. However, extrapolating the 0.1 Gyr time scale to
reach steady state in our lowest turbulent diffusion coefficient to 20
times lower diffusion would give a time scale of 2 Gyr.  This
computation may hence not have been carried out over a long enough
time to probe if this front will eventually reach a steady state or
not.
 
 The multifluid nature of our computations is not essential to the
present study. Nevertheless, it allows us to uncover a potentially
interesting partial pressure effect.  A slight neutral/charges
velocity drift of the order of 0.01~km.s$^{-1}$ persists in the
hot phase throughout the simulation. This drift is sustained by the
partial pressure equilibration. Indeed, photo-ionisation is more
effective in the hot phase and a partial pressure gradient drives a
flow of charged ions towards the cold medium. This is the cause of a
current of charges from the hot toward the cold medium : C$^+$ ions
accumulate at the surface of the cloud where their radiative cooling
produces a dip in the temperature profile (see figure \ref{Tn}).  If
self-gravity of the cloud was taken into account, this would produce a
convective instability at the surface of the cloud that would soon
smooth out the temperature and C$^+$ profiles. This C$^+$ effect might
also be responsible for the failure of the front to reach a completely
static state.

\subsection{Effect of turbulent diffusion}

  Although we display the results for big cloud simulations, we
restrict our analysis to small cloud simulations in this section. The big
cloud simulation was run following the same numerical experiment as
for the small cloud, but we investigated only two different values for
the turbulent diffusion coefficient. Big cloud simulations are discussed
in section \ref{bigcloud}.

  We restart our computation from $t=0.1$~Gyr in the reference run and
switch on turbulent diffusion while resetting the time to 0. After
about another 0.1~Gyr has occurred, a static state is
reached. The time to reach a steady state is much shorter than in
the case without diffusion, because the effective diffusion time
scales are much smaller. The ion current is still present, but the
diffusion smoothes out the dip in temperature and in C$^+$
composition. This seems to allow the front to find a final
equilibrium. The pressure in this final equilibrium turns out to be
nearly 40\% less than in the case without turbulent diffusion at time
0.1~Gyr.

  We now probe the dependence of this steady state on the magnitude of
the turbulent diffusion coefficient. We compare the state of the
simulation box at time 0.1~Gyr for different turbulent length scales
$L$ ranging from $1.25\times 10^{-4}R$ to $1.6\times 10^{-2}R$ with a
factor 2 increment. With $R=5$~pc, this range extends from
$L=6$~10$^{-4}$~pc to $L=0.08$~pc. Note that the mean free path is at
a scale of around $3 \times 10^{-5}$~pc in the cold medium, which
corresponds to $6~\times~10^{-6} R$ and makes molecular diffusion
negligible by at least a factor 20 with respect to turbulent diffusion
in all our computations. The upper bound $L=0.08$~pc is roughly a
seventh of the size of the cold cloud at this level of diffusion
(remember $L$ is set as a fraction of the size of the computational
box but it is difficult to predict what the actual size of the cloud
will be before actually running the simulation). As a point of
comparison, \cite{X95} built their diffusion coefficients with
$L=0.1-0.5$~pc and a fixed velocity of 1 km.s$^{-1}$ for a molecular
cloud of size 1~pc.

\begin{figure}
\centerline{
\includegraphics[width=7cm,angle=-90]{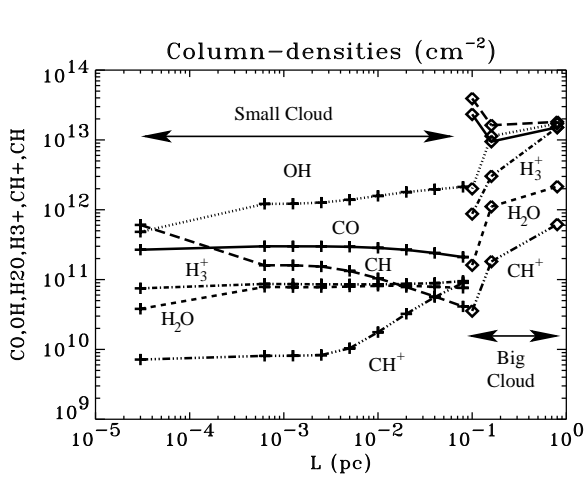}
} 
 \caption{CO (solid), OH (dotted), H$_2$O (dashed), H$_3^+$
(dash-dotted), CH$^+$ (dash-dot-dot-dot) and CH (long-dash)
column-densities (cm$^{-2}$) integrated over the computational box for
condensation fronts at time 0.1~Gyr with various diffusion lengths $L$
(the big cloud run with the highest $L$ is actually displayed at time
0.02~Gyr, when the WNM has disappeared).  Plus signs are for the
small cloud and diamonds stand for the big cloud runs.  Note that the
values for the runs without turbulent diffusion at time 0.1~Gyr have
been arbitrarily placed at $L=3 \times 10^{-5}$~pc (small cloud) and
$L=0.1$~pc (big cloud).  }
\label{NL}
\end{figure}

 Figure \ref{NL} displays the total column-densities of the molecular
species of main interest for various diffusion coefficients. Except
for the column-densities of CH which decreases, and those of CO and
H$_3$$^+$ which stay at the same level, the effect of turbulent
diffusion is to boost the production of molecular species. In the
following, we proceed to a more detailed investigation of the
structure of the front:  for each molecule of interest in each
simulation, we computed which source and sink terms were dominant,
including reaction and diffusion effects. This reveals that H$_2$
molecules mixed into hot gas are the main cause for the
diffusion-enhanced production of molecules.

\subsubsection{Thermal and density structure}

\begin{figure}
\centerline{
\begin{tabular}{c}
\includegraphics[width=7cm,angle=+90]{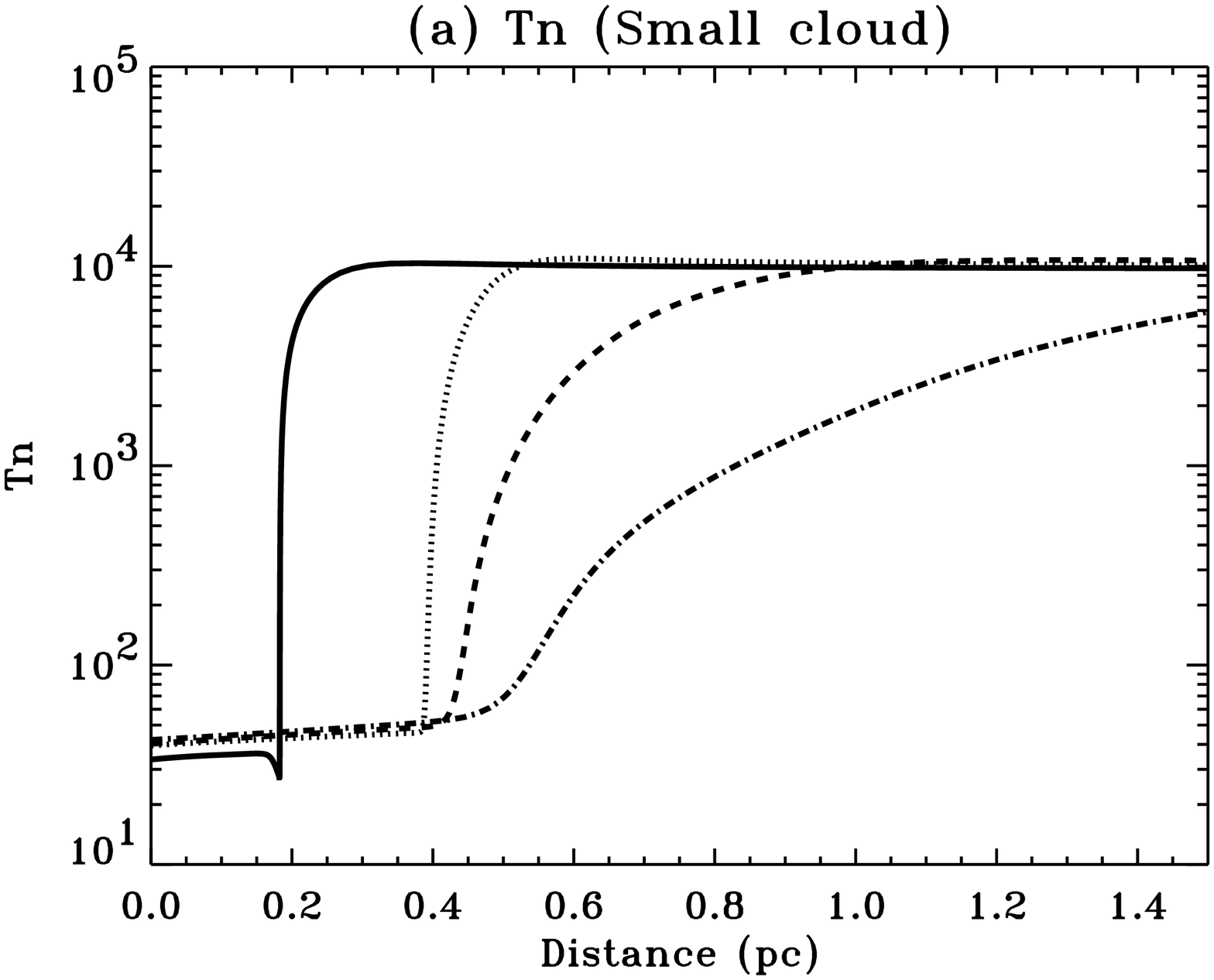}\\
\includegraphics[width=7cm,angle=+90]{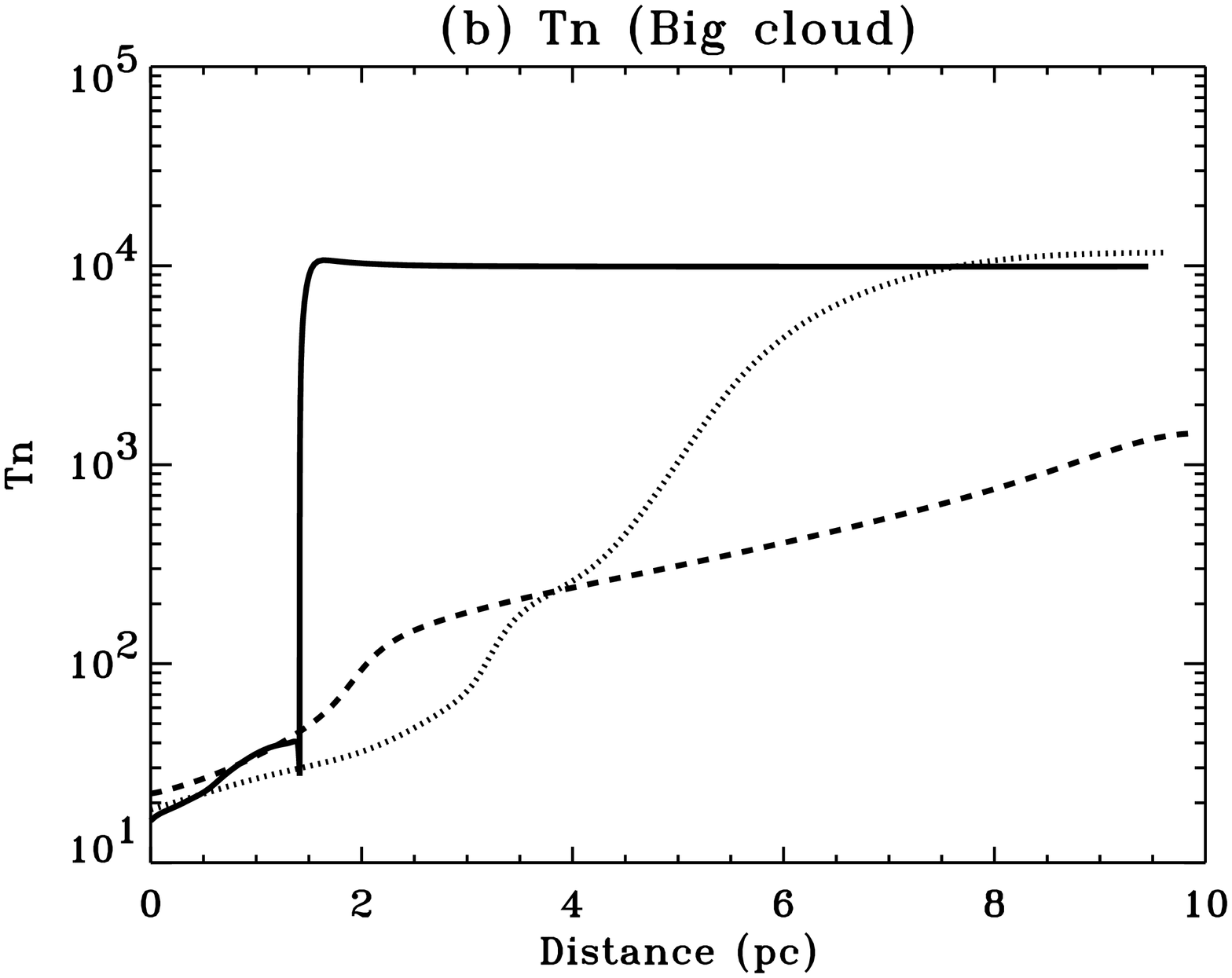}
\end{tabular}
} 
\caption{ Temperature (K) structure of the front at steady-state for various
  diffusion coefficients:
{\it (a) upper panel} (standard run) the solid line is for
molecular diffusion, the dotted line for $L=6 \times 10^{-4}$~pc, the
dashed line for $L=10^{-2}$~pc and the dash-dotted line for 
$L=8 \times 10^{-2}$~pc ;
{\it (b) lower panel} (big cloud run) the solid line is for
molecular diffusion, the dotted line for $L=0.16$~pc and the dashed
line for $L=0.8$~pc.
}
\label{Tn}
\end{figure}

 As illustrated in figure \ref{Tn}, the main effect of increasing the
diffusion is to spread out the length of the front. As a result of the
heat exchange between the hot and the cold phase, the cold phase becomes
slightly hotter and the hot phase becomes slightly colder. 

\begin{figure}
\centerline{
\begin{tabular}{c}
\includegraphics[width=7cm,angle=+90]{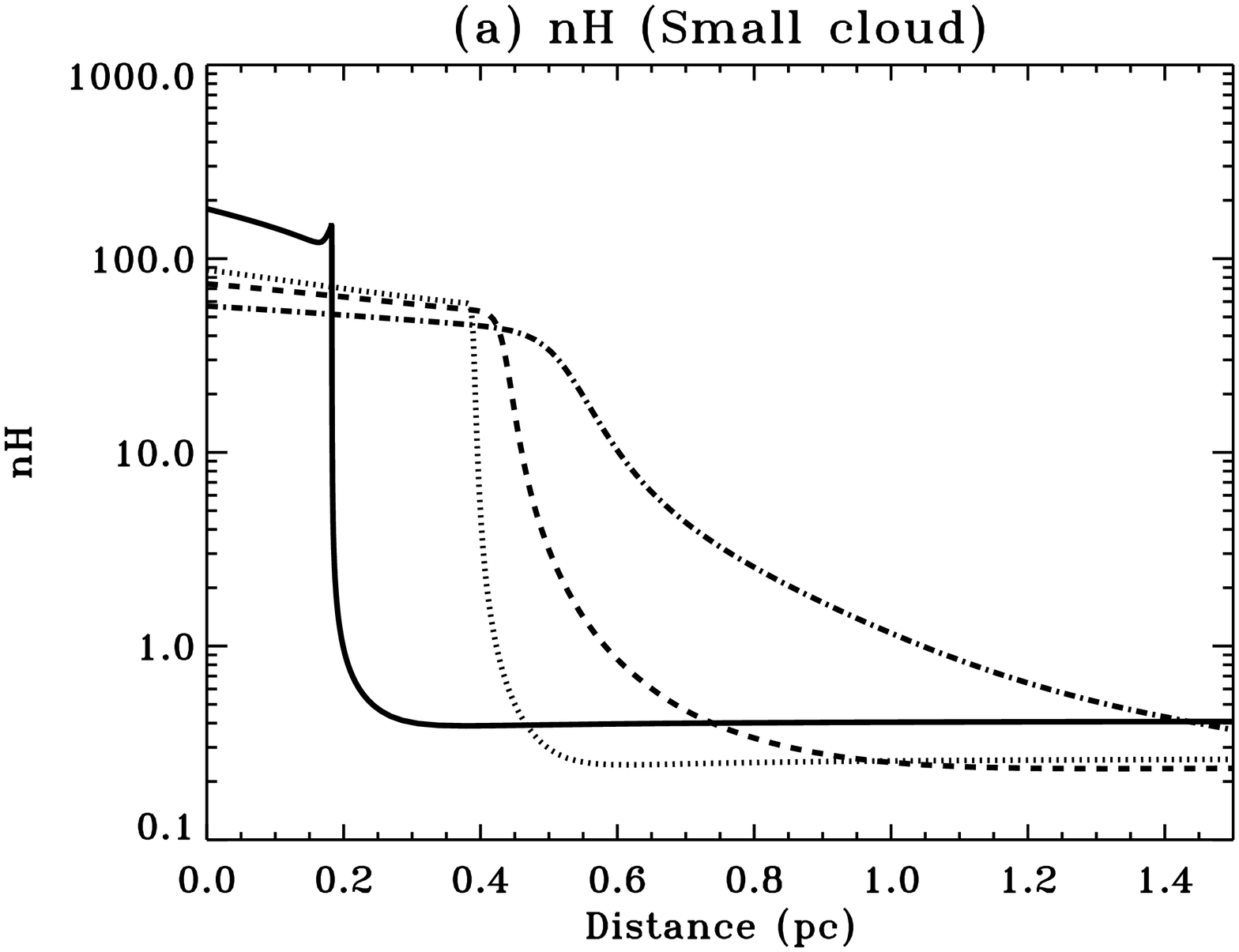}\\
\includegraphics[width=7cm,angle=+90]{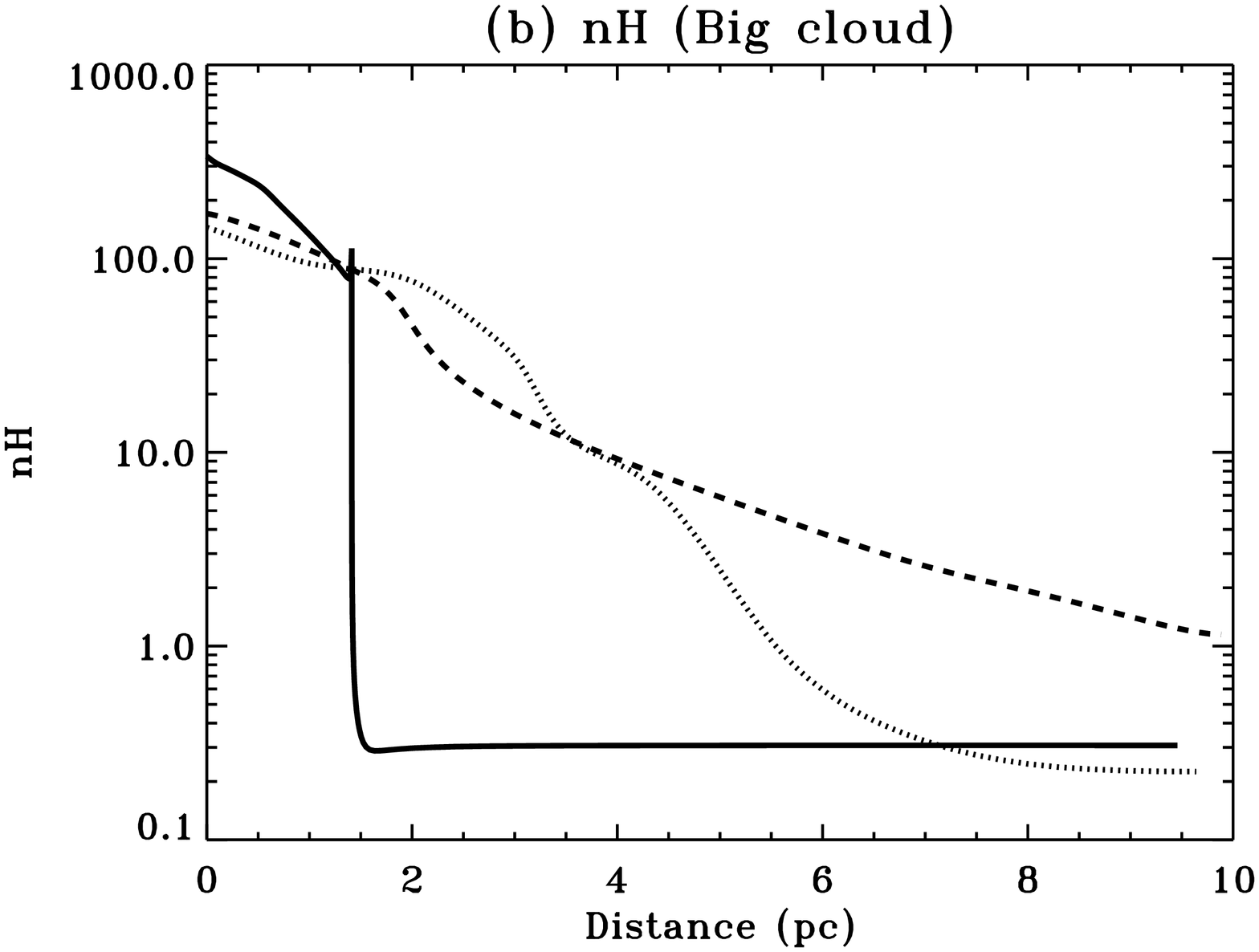}
\end{tabular}
} 
\caption{Density (cm$^{-3}$) structure of the front for various
diffusion coefficients (same labels as figure \ref{Tn}).
}
\label{nH}
\end{figure}

  As already mentioned, the total pressure at steady state decreases
when diffusion is enhanced. Besides, the cold medium temperature increases
slighlty in the case with diffusion.  As a result, the total density
in the cloud decreases by more than a factor 2 in the case with
diffusion (see figure \ref{nH}).  The cloud expands to keep its total
column-density constant (remember that the total mass of the
simulation box remains constant).

 Everywhere in all simulations the dominant heating process is
  the photo-electric effect. The main cooling processes are from the
  inside of the cloud to the WNM (left to right): C$^+$ cooling,
  diffusive cooling (on the WNM side of the diffusion front) and H
  (Lyman $\alpha$) cooling. In the WNM, O cooling is also important,
  but amounts to at most a half of the Lyman $\alpha$ cooling. When
  turbulent diffusion is present, molecules are abundant in the WNM
  (see below) and H$_2$ cooling sets in in between the region where
  diffusive cooling and Lyman $\alpha$ cooling prevail. Note that
  although O collisional excitation by electrons is adopted, we use
  H$_2$ cooling functions from \cite{L99} which include excitation by
  H, He and H$_2$ only.

\subsubsection{H$_2$ abundance}

  The abundance of H$_2$ usually results from the balance
between its formation on grains and its destruction by
photodissociation.  However, for cases with turbulent diffusion, in
the intermediate region between the CNM and the WNM, the main source
of H$_2$ is the divergence of its diffusive flux.

\begin{figure}
\centerline{
\begin{tabular}{c}
\includegraphics[width=7cm,angle=+90]{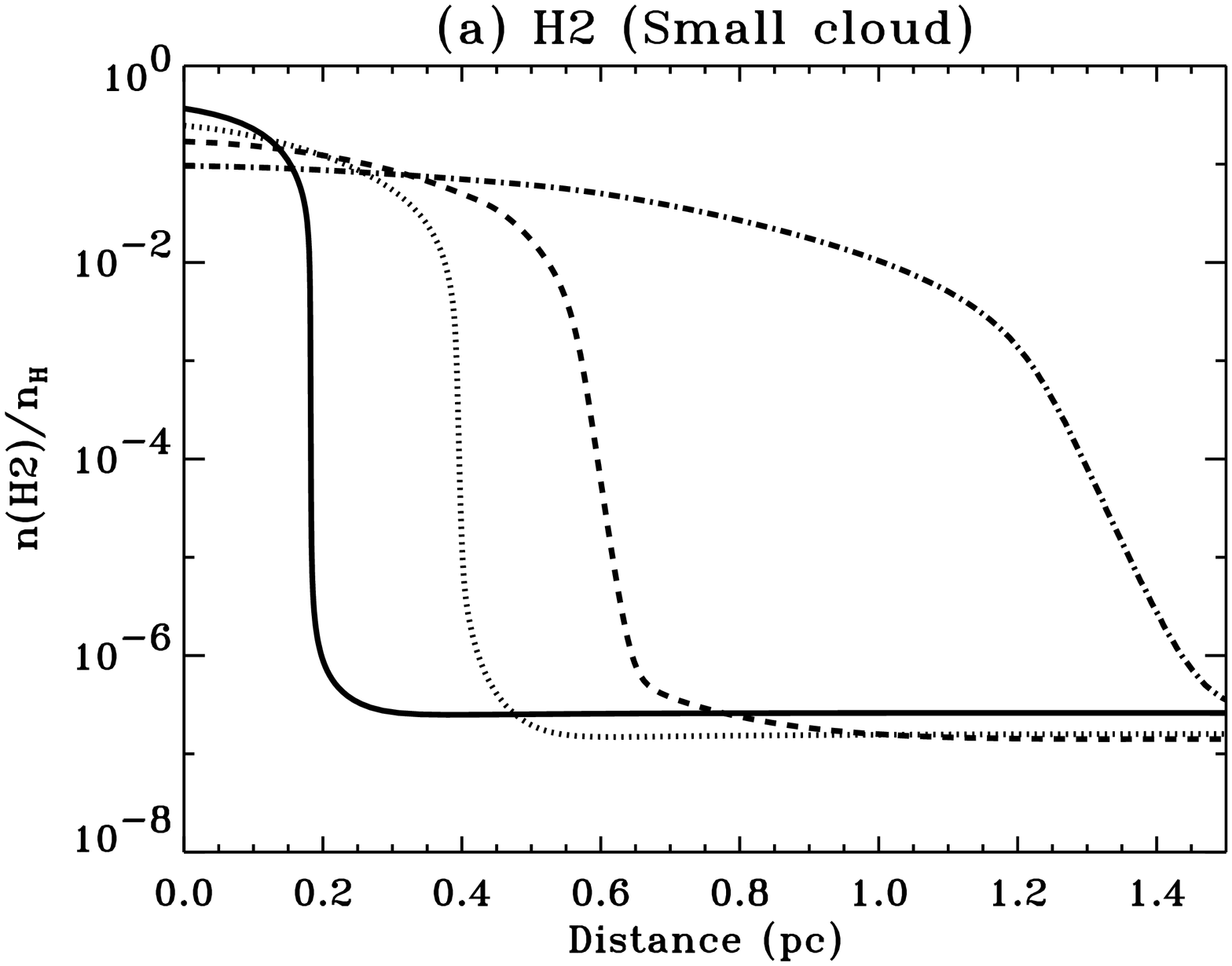}\\
\includegraphics[width=7cm,angle=+90]{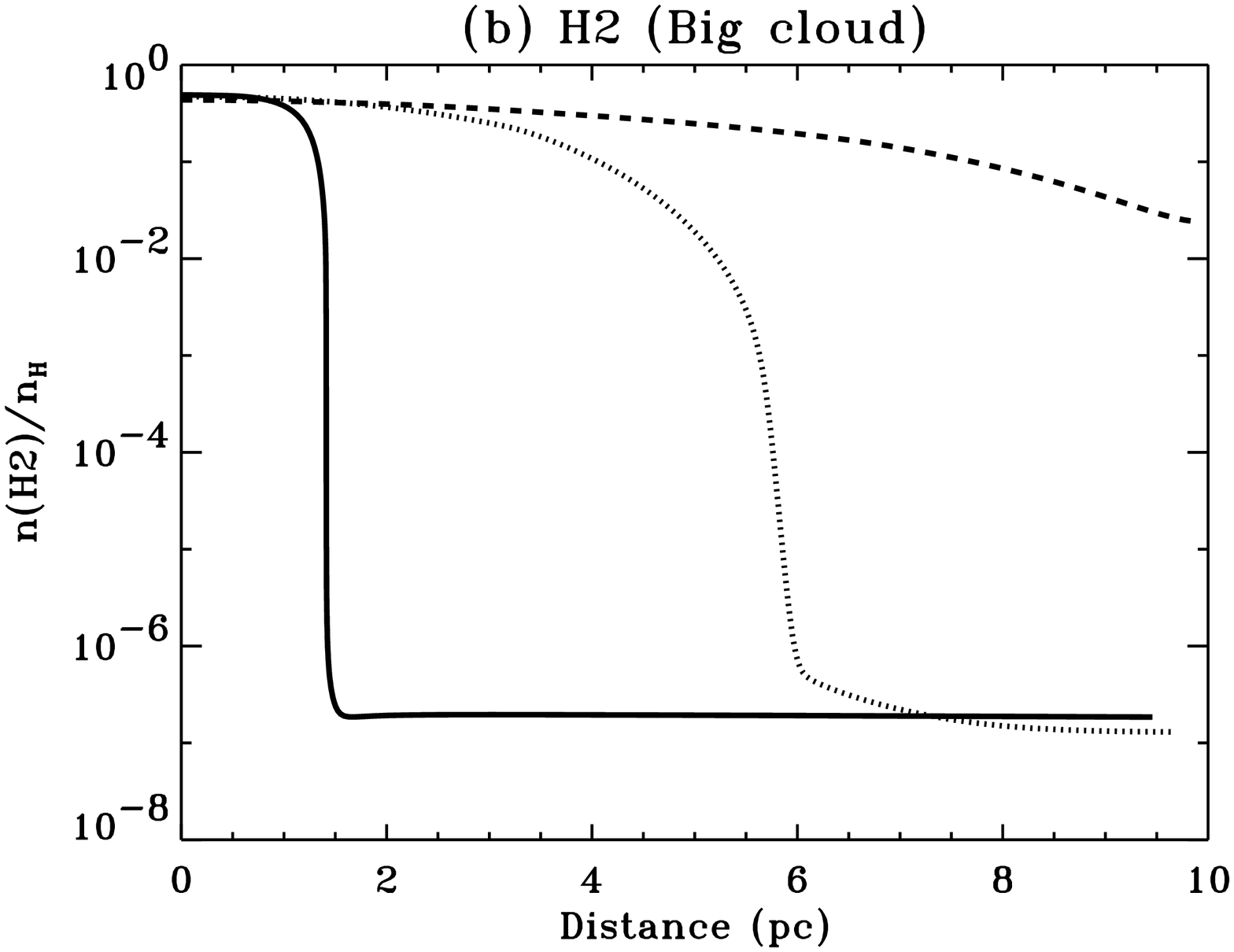}
\end{tabular}
} 
\caption{Abundance profile of H$_2$ for various diffusion
coefficients (same labels as figure \ref{Tn}).  }
\label{H2}
\end{figure}

An increase in the diffusion coefficient considerably enhances the
abundance of H$_2$ in the diffusion front (see figure \ref{H2}). This
is mainly a diffusion effect: diffusion mixes the cold molecular gas
from the cloud with the external hot atomic gas. The diffusion flux
provides a source of H$_2$ molecules in the hot phase which is strong
enough to counteract photo-dissociation and manages to sustain high
concentration levels of H$_2$ in a hot medium.  Hence turbulence
brings H$_2$ molecules in contact with hotter gas: this effect will be the
source of a lot of efficient molecular chemistry.

  The decrease in H$_2$ abundance in the cloud for enhanced diffusion
results from the combined effects of the general spread in H$_2$
(diffusion pumps H$_2$ from the cold cloud) and the lower density
(H$_2$ formation on grains is less efficient).

\subsubsection{CH$^+$ abundance}

The most spectacular consequence of increasing the temperature of
H$_2$ rich gas is the production of the CH$^+$ molecular ion.  In
absence of diffusion, CH$^+$ is produced via the reaction
C$^+$~+~H~$\rightarrow$~CH$^+$ and destroyed by photodissociation in
the hot gas or by CH$^+$~+~H~$\rightarrow$~C$^+$~+~H$_2$ in the cold
gas.

\begin{figure}
\centerline{
\begin{tabular}{c}
\includegraphics[width=7cm,angle=+90]{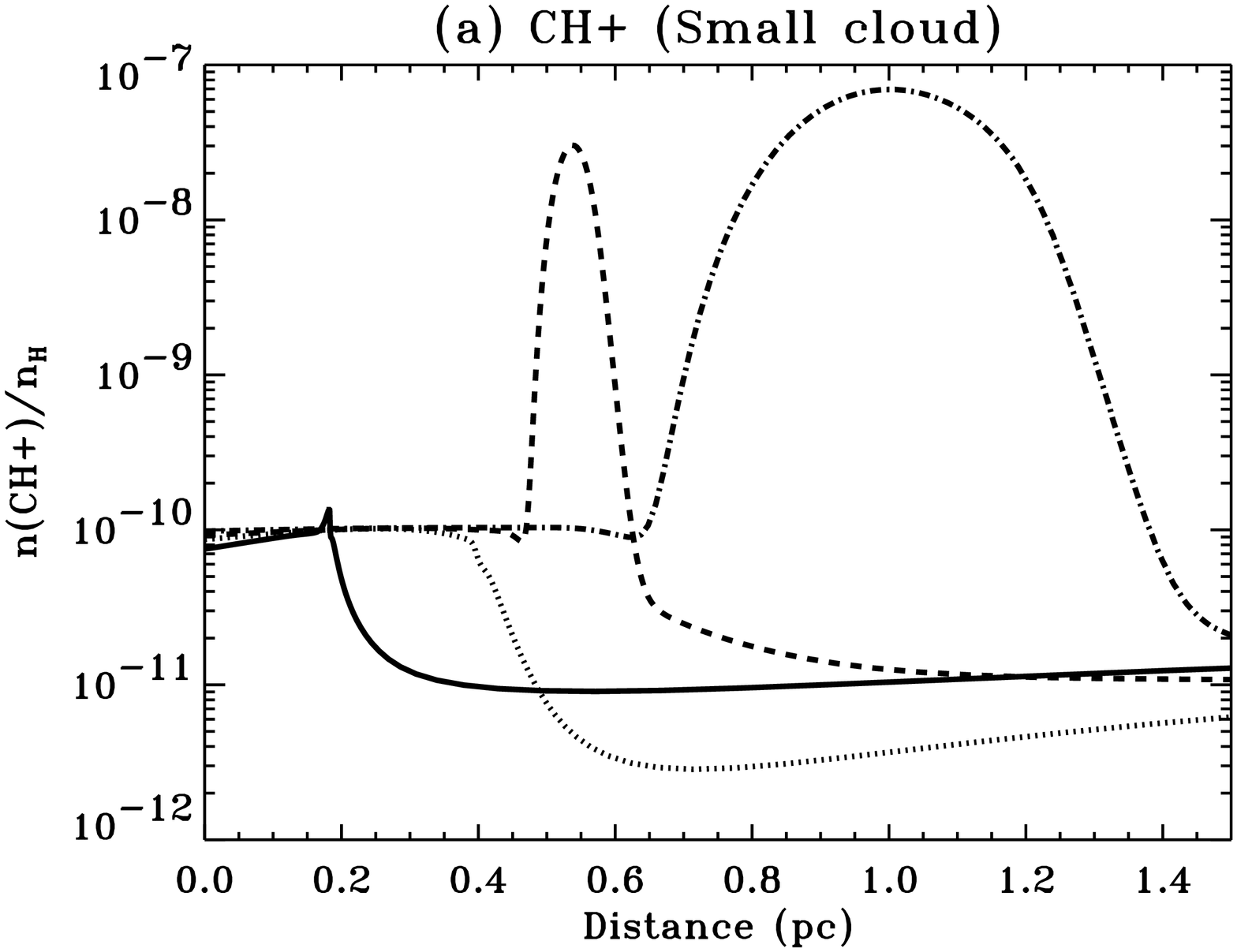}\\
\includegraphics[width=7cm,angle=+90]{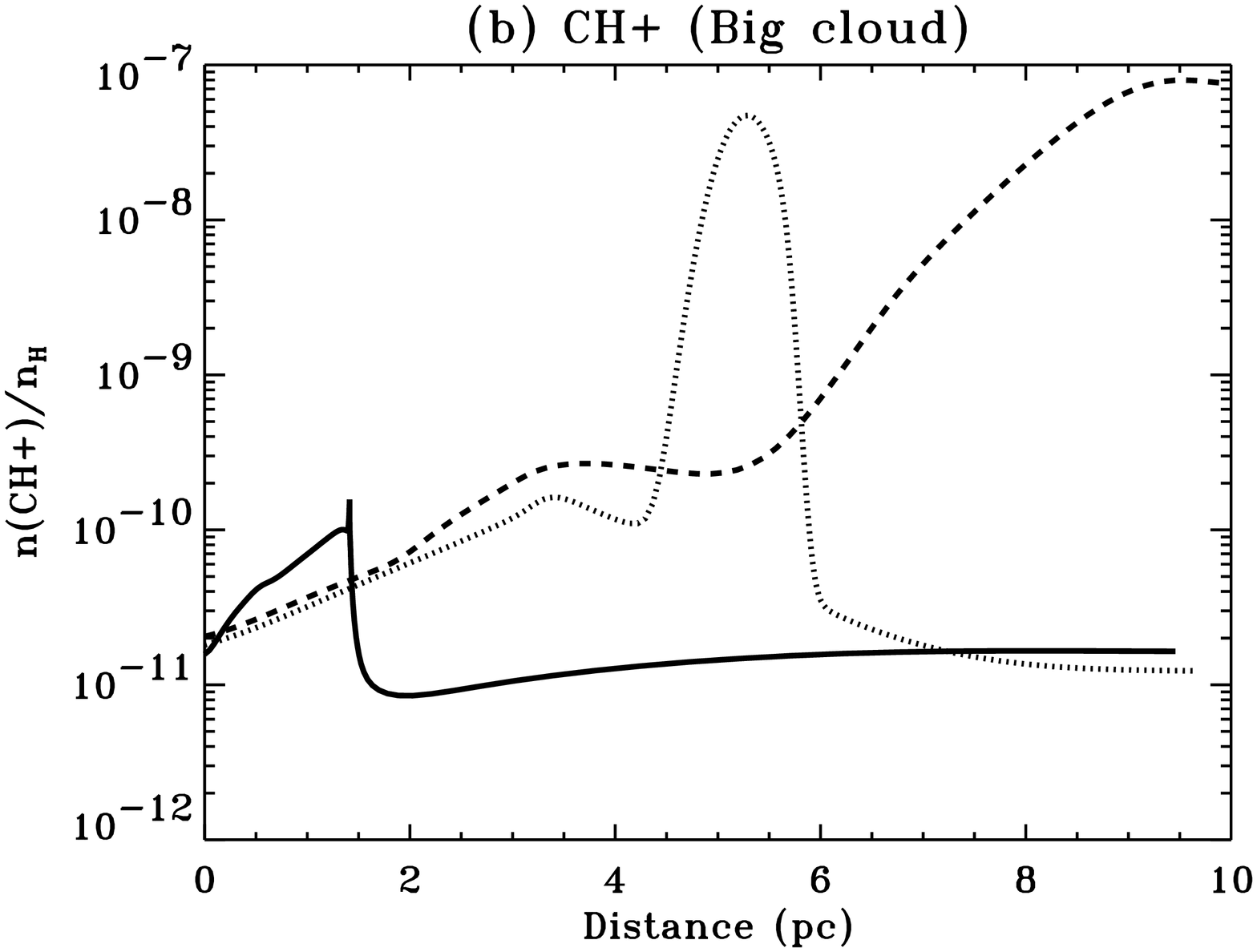}
\end{tabular}
} 
\caption{Abundance profile of CH$^+$ for various
diffusion coefficients (same labels as figure \ref{Tn}).  }
\label{CH+}
\end{figure}

 However, the reaction C$^+$~+~H$_2$~$\rightarrow$~CH$^+$~+~H can be a
much more efficient source for CH$^+$ provided that the 4640~K
activation temperature is overcome in presence of H$_2$.  As clearly
seen in figure \ref{CH+}, diffusion provides an excellent means of
bringing together high temperature and H$_2$ rich gas in order to
make CH$^+$ through this channel.


\subsubsection{H$_3$$^+$ abundance}

\begin{figure}
\centerline{
\begin{tabular}{c}
\includegraphics[width=7cm,angle=+90]{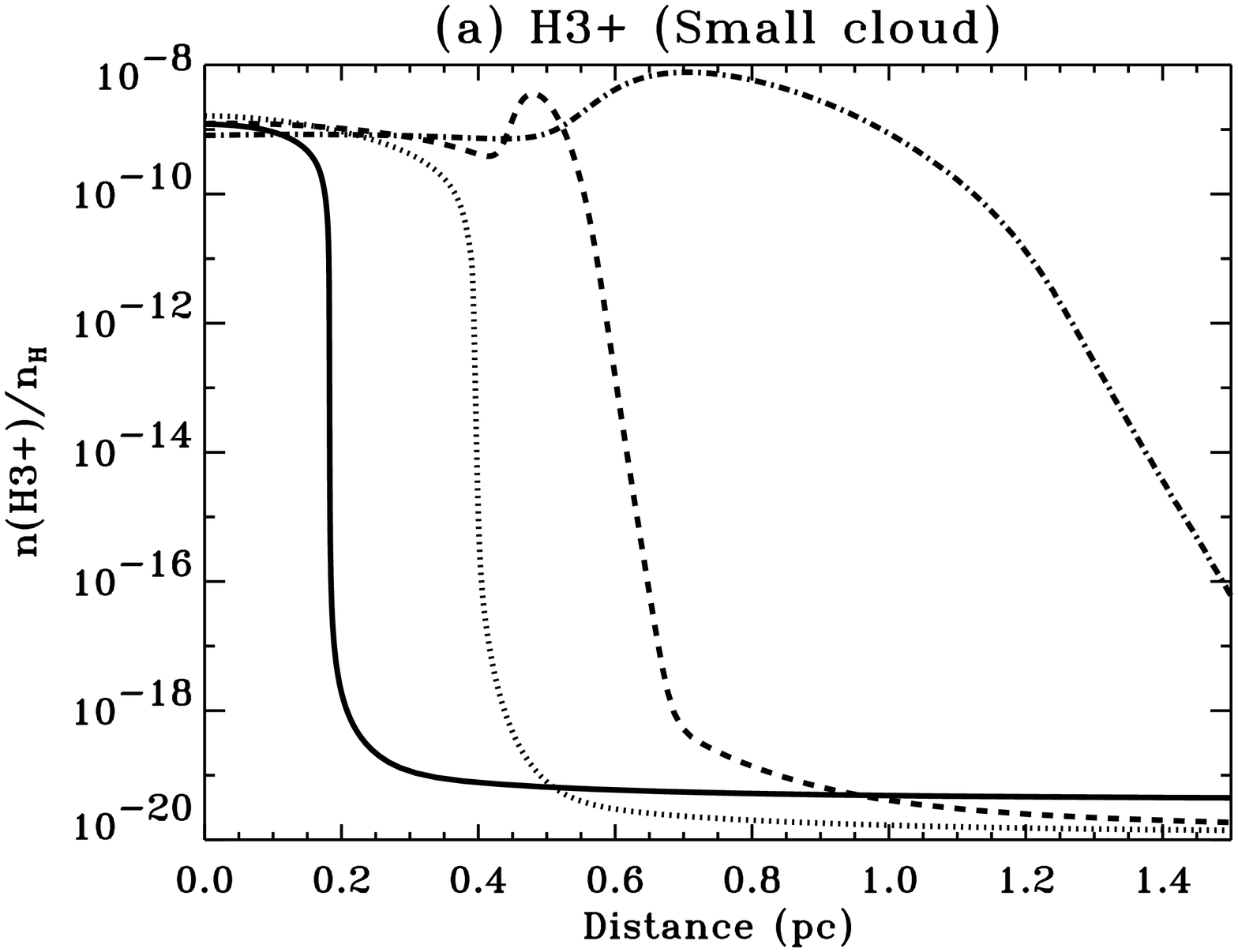}\\
\includegraphics[width=7cm,angle=+90]{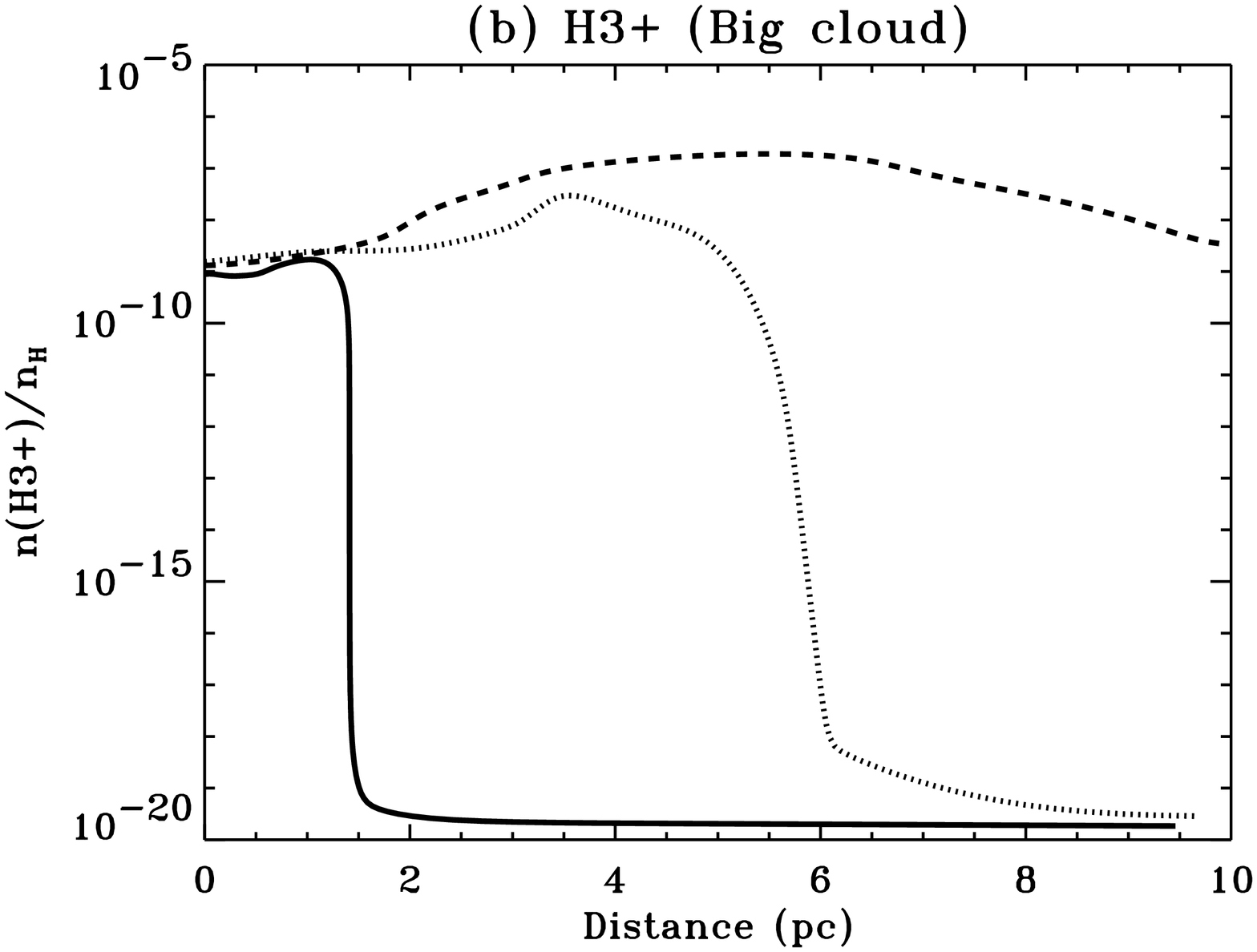}
\end{tabular}
} 
\caption{Abundance profile of H$_3$$^+$ for various
diffusion coefficients (same labels as figure \ref{Tn}).  }
\label{H3+}
\end{figure}

H$_3$$^+$ is made via the reaction
H$_2$$^+$~+~H$_2$~$\rightarrow$~H$_3$$^+$~+~H and destroyed through
dissociative recombination with electrons. As a result, $n($H$_3^+)=
{\zeta n(H_2)}/ {k_e n(e)}$ remains valid throughout most of the
region where H$_3^+$ is present. The recombination rate $k_e$ is a
decreasing power of the temperature (it behaves like
$1/\sqrt{T}$). Furthermore, it turns out that the electron abundance
$n(e)$ also scales approximately like $\sqrt{T}$ in the intermediate region.
One should therefore expect more H$_3$$^+$ in the hot phase. However,
photo-dissociation usually destroys the source of H$_2$ molecules in
the hot phase. Turbulent diffusion allows to replenish part of the
high temperature gas in H$_2$ molecules. This effect produces a peak
of H$_3$$^+$ inside the diffusion front as illustrated in figure
\ref{H3+}.

\subsubsection{OH abundance}

\begin{figure}
\centerline{
\begin{tabular}{c}
\includegraphics[width=7cm,angle=+90]{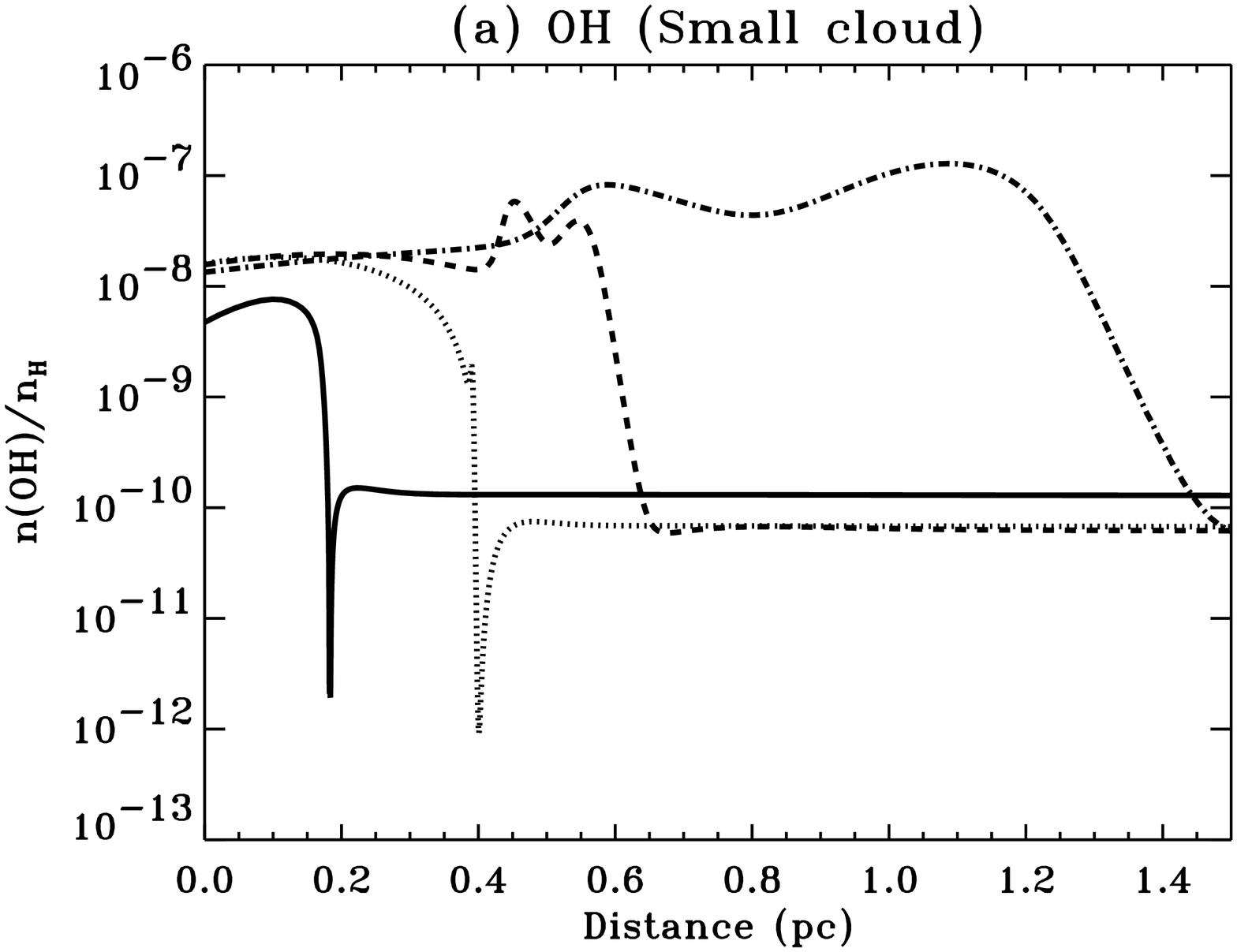}\\
\includegraphics[width=7cm,angle=+90]{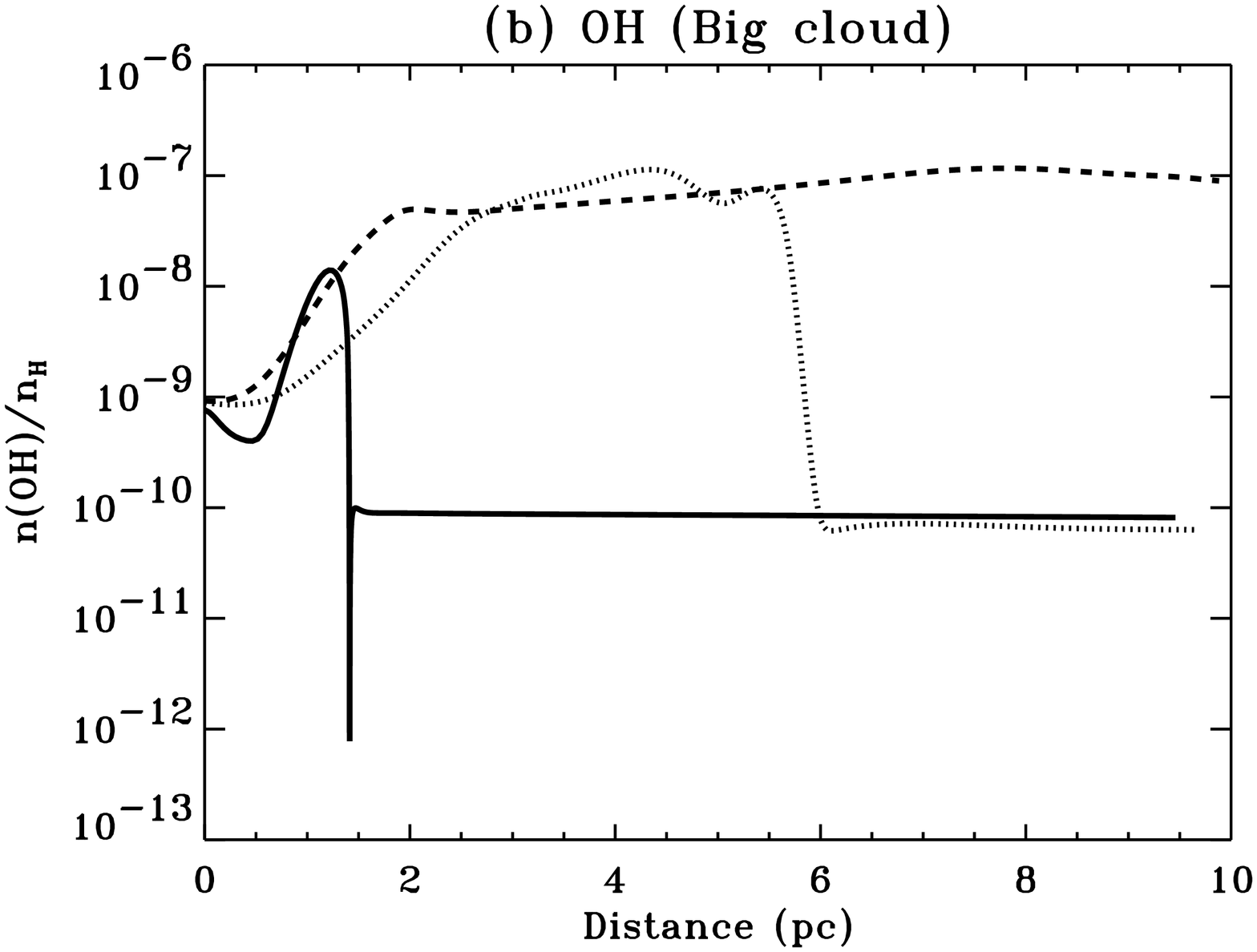}
\end{tabular}
} 
\caption{Abundance profile of OH for various diffusion
coefficients (same labels as figure \ref{Tn}).  }
\label{OH}
\end{figure}

In our simulations, OH usually results from the competition between
the recombination of H$_2$O$^+$ and photo-dissociation (for the
conditions where the H$_2$ fraction is greater than about 0.2-0.3,
H$_2$O$^+$ is destroyed by H$_2$ and recombination of H$_3$O$^+$
becomes the main source, Pineau des For\^ets 2007, private
communication). However, the H-atom transfer
O~+~H$_2$~$\rightarrow$~OH~+~H can become the dominant source if its
2980~K activation temperature is overcome in presence of H$_2$. Figure
\ref{OH} shows that this is indeed the case at the CNM/WNM interface
when turbulent diffusion is able to bring enough H$_2$ at sufficiently
high temperature.

An other noticeable effect of turbulent diffusion is a factor 3 to 5
rise of the relative abundance of OH in the cold cloud from the case
without turbulent diffusion to the case with turbulent diffusion. This
is less straightforward to understand.  Indeed diffusion slightly
increases the temperature and strongly decreases the density in the
cold cloud which is expected to decrease the recombination rate of
H$_2$O$^+$, hence lowering the OH abundance.  However, H$_2$O$^+$
comes from H$^+$ via the reaction
H$^+$~+~O~$\rightarrow$~O$^+$~+~H followed by
OH$^+$~+~H$_2$~$\rightarrow$~H$_2$O$^+$~+~H. The former reaction is
subject to a 227~K activation temperature, which rate increases by at
least a factor 6 in the cold cloud between the case with and without
diffusion. H$^+$ is itself subject to at least a factor 3 rise in
relative abundance in the cold medium due to its lower recombination
rate thanks to a slightly higher temperature and a less dense gas (the
electron fraction remains roughly the same, as it is bound to the
abundance of C$^+$).  The net effect is a rise in OH relative
abundance due to the change in the physical conditions inside the CNM.

The very sharp dip in OH relative abundance at the surface of the
cloud when turbulent diffusion is switched off (see the solid line on figure
\ref{OH}) is in fact composed of two dips.  The inner one is caused by
the dip in temperature and the sensitivity of the H$^+$~+~O reaction
rate to temperature. The outer one is caused by the decrease of H$_2$
that lowers the H$_2$O$^+$ rate of production in a region where
O~+~H$_2$ is not activated yet. Only the outer dip remains for
$L=6~\times~10^{-4}$~pc (see the dotted line on figure \ref{OH}).

\subsubsection{CO abundance}

 In the densest parts of the cloud, CO is produced through the
reaction HCO$^+$~+~e$^-$~$\rightarrow$~CO~+~H. That reaction accounts
for the inner 0.1~pc in the small cloud cases and the inner 3~pc for
the big cloud with the highest turbulent diffusion coefficient (in the
latter case, CH~+O~$\rightarrow$~CO~+~H also prevails in the inner
0.5~pc). HCO$^+$ is indeed pointed out as one of the most likely sources
for CO in diffuse and translucent clouds \citep{LL98}.

\begin{figure}
\centerline{
\begin{tabular}{c}
\includegraphics[width=7cm,angle=+90]{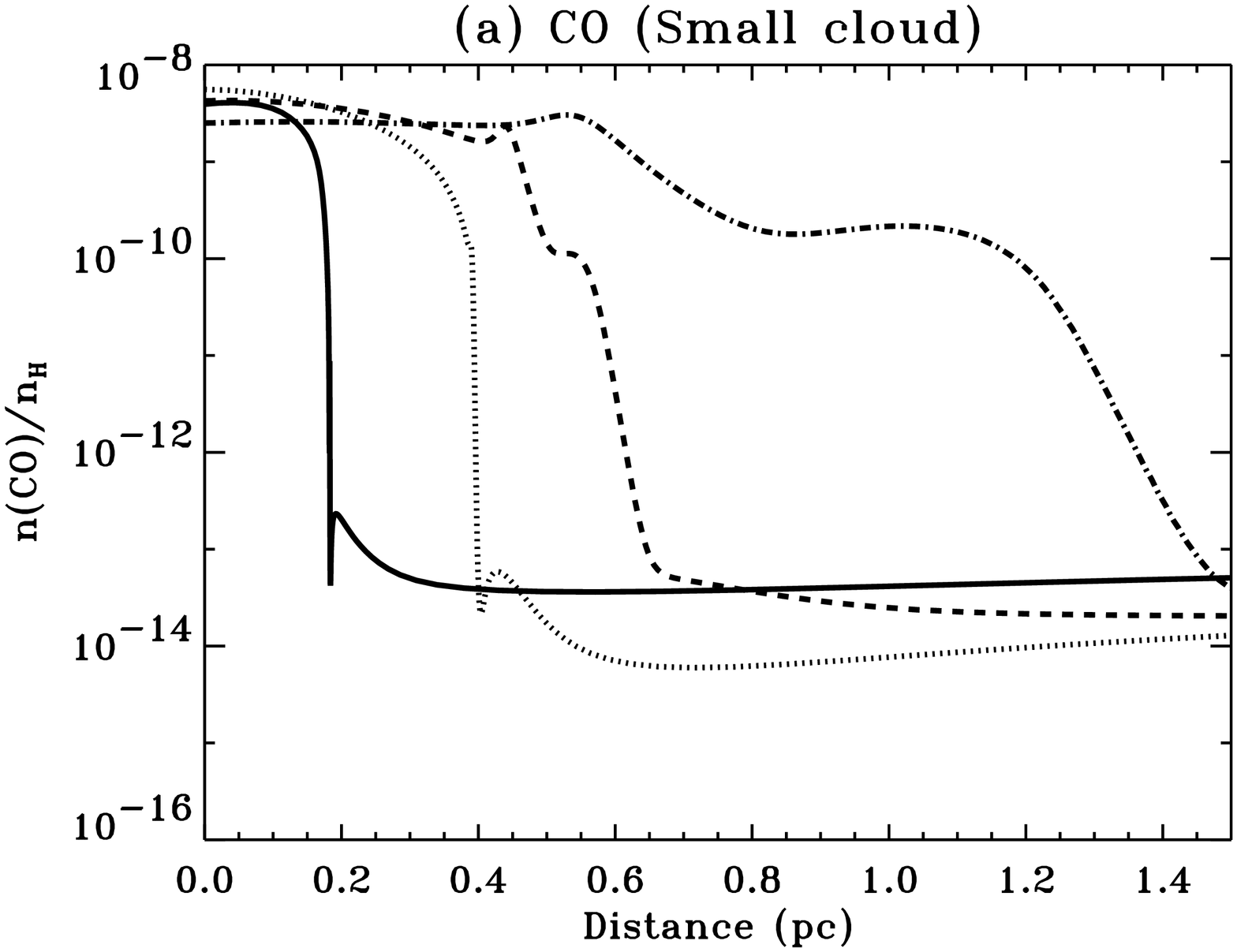}\\
\includegraphics[width=7cm,angle=+90]{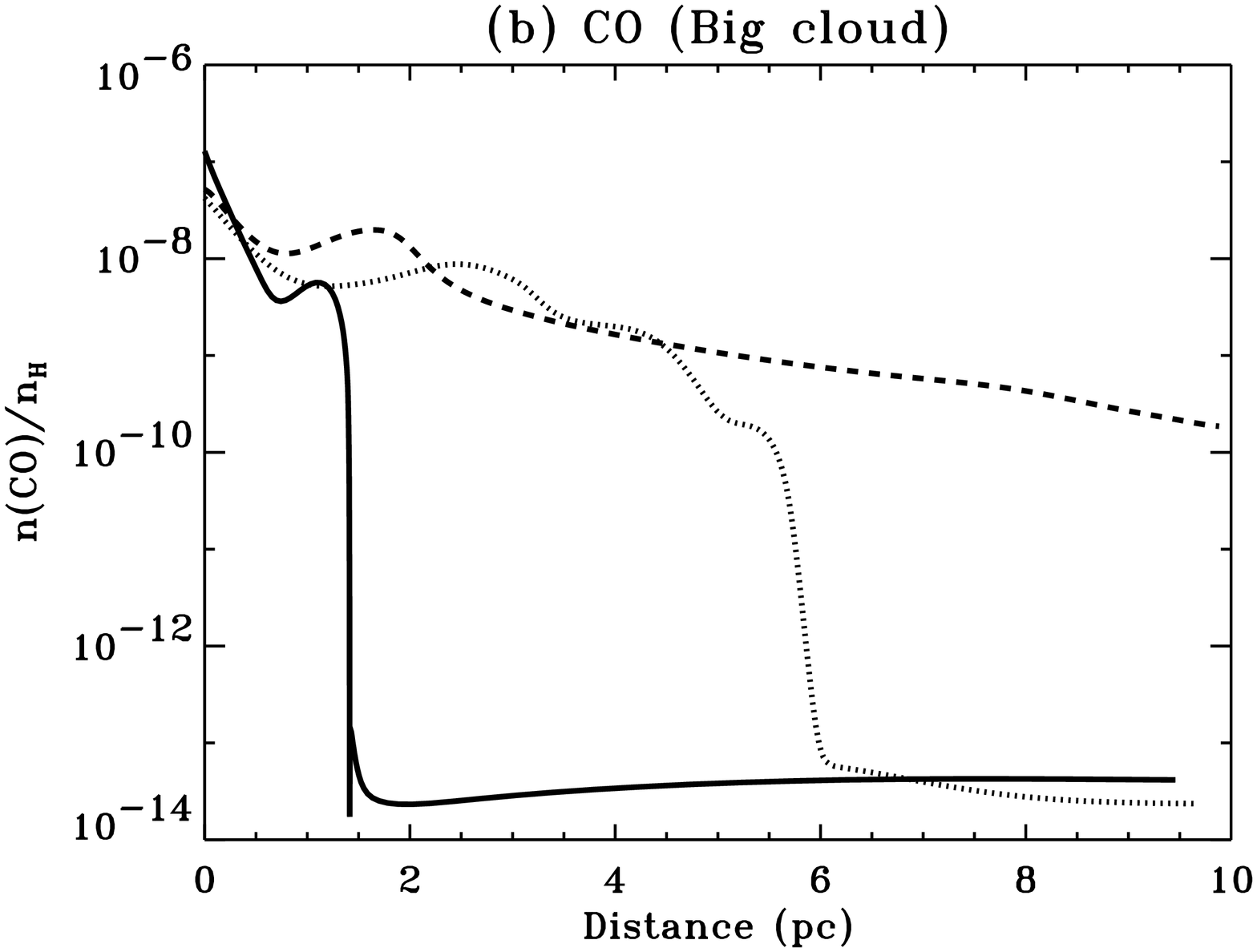}
\end{tabular}
} 
\caption{Abundance profile of CO for various diffusion
coefficients (same labels as figure \ref{Tn}).  }
\label{CO}
\end{figure}

 In the rest of the simulation box, the molecule CO is most
 often produced by C$^+$~+~OH~$\rightarrow$~CO~+~H$^+$ and destroyed
 by photo-dissociation.  In these regions its relative abundance is
 hence directly linked to the relative abundance of OH and density, as
 demonstrated by figure \ref{CO}. In a small region at the right
 hand side of the diffusion front, the reaction
 CO$^+$~+~H~$\rightarrow$~CO~+~H$^+$ is the most efficient source of
 CO. For high diffusion coefficients, the abundance profile of CO can
 also be subject to diffusive mixing that removes CO from the CNM.  The
 enhanced CO in the front, due to enhanced OH, is never strong enough
 to produce a significant local maximum in the diffusion front.

\subsubsection{H$_2$O abundance}

\begin{figure}
\centerline{
\begin{tabular}{c}
\includegraphics[width=7cm,angle=+90]{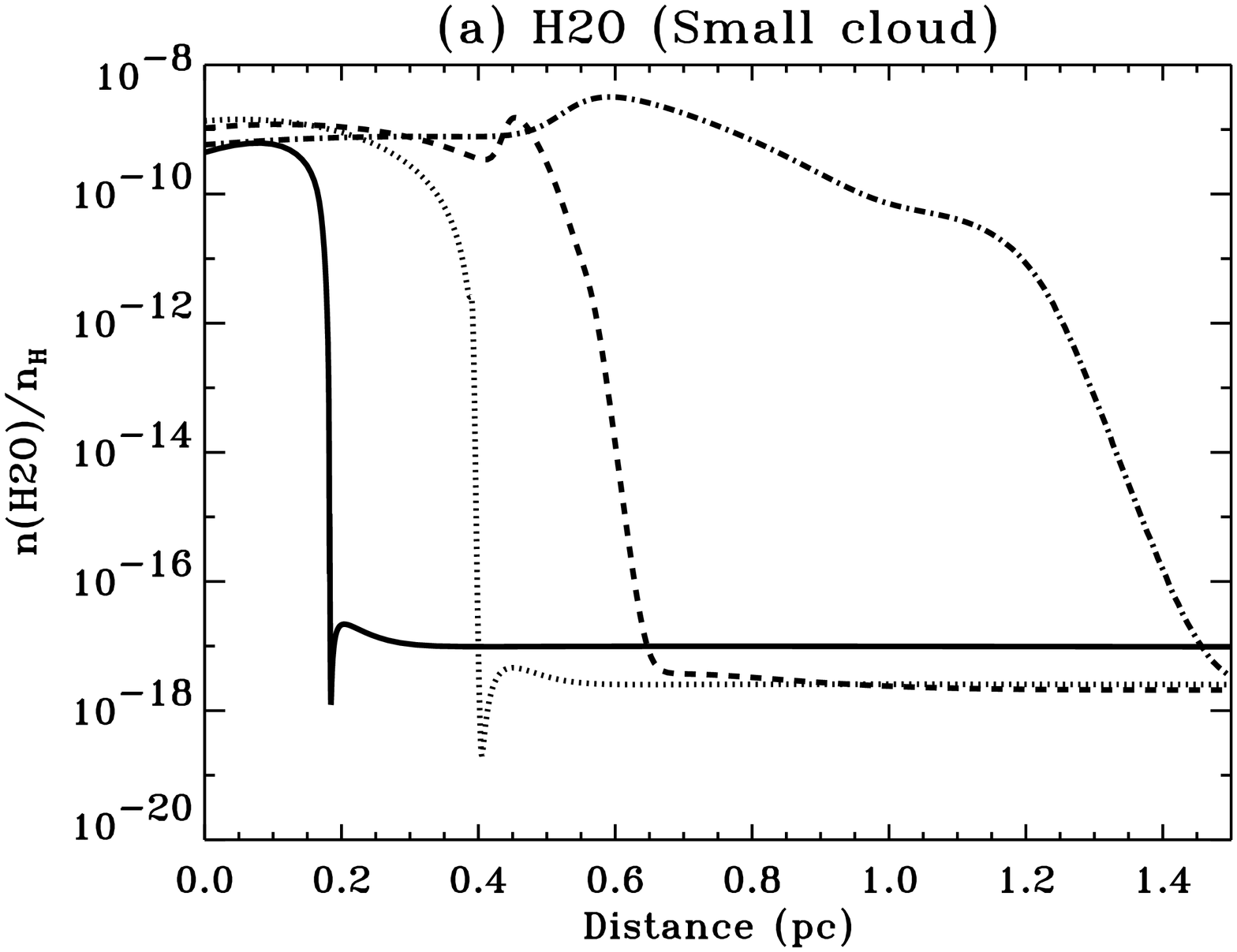}\\
\includegraphics[width=7cm,angle=+90]{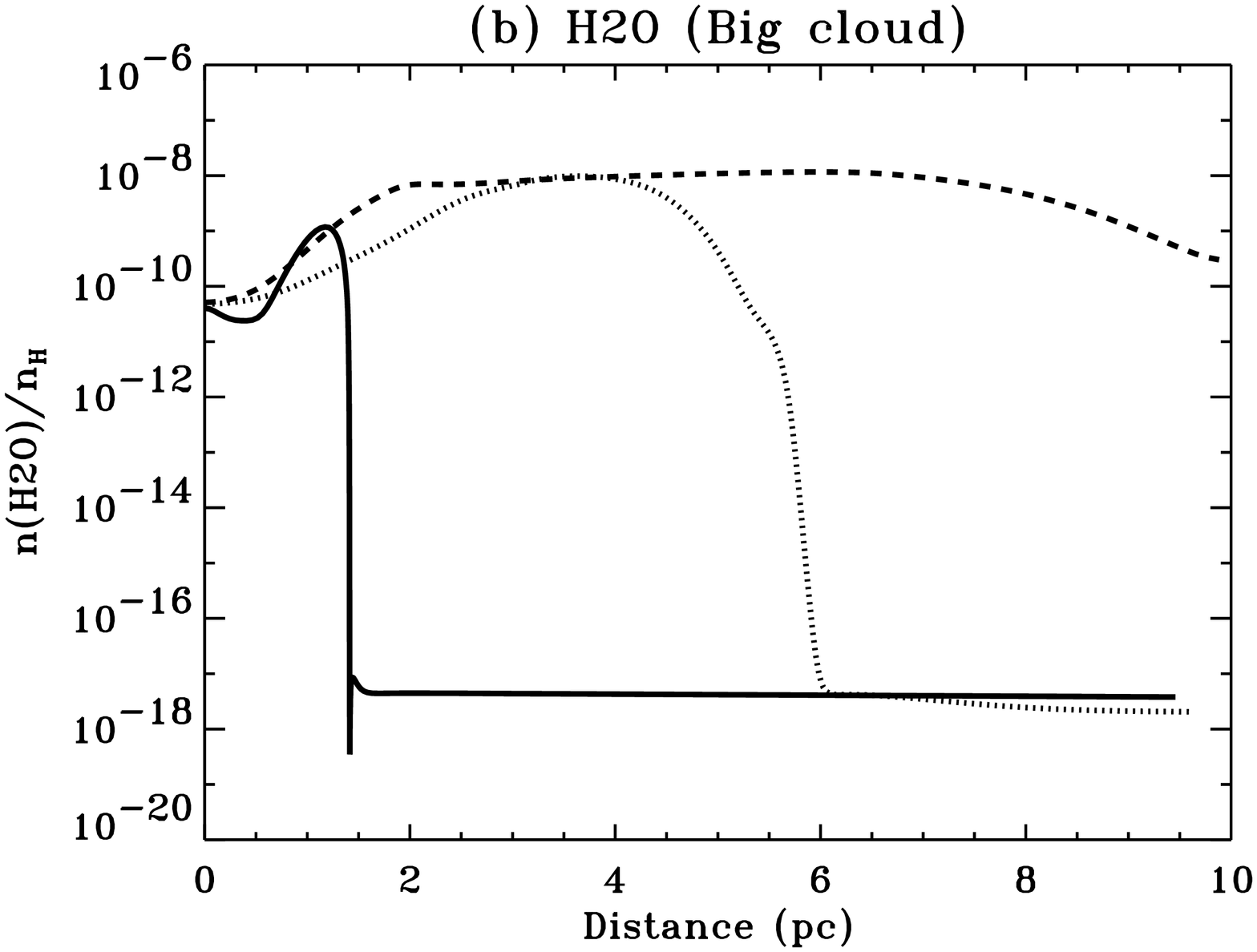}
\end{tabular}
} 
\caption{Abundance profile of H$_2$O for various
diffusion coefficients (same labels as figure \ref{Tn}).  }
\label{H2O}
\end{figure}

H$_2$O is generally produced by the recombination of the molecular ion
H$_3$O$^+$, which comes from proton exchange between H$_2$O$^+$ and
H$_2$ and is hence linked to the abundance of H$^+$ as is OH in the
cold cloud. H$_2$O experiences the same rise in abundance as OH in the
cold cloud (see figure \ref{H2O}) but has an extra dependence on the
H$_2$ fraction. This effect might be responsible for the actual decrease of
the H$_2$O relative abundance in the CNM at high diffusion
coefficients.  However, the dominant source of H$_2$O {\it in the
diffusion front} is proton exchange between H$_2$ and OH, as both
molecules are abundant in the front for the reasons stated above.

\subsubsection{CH abundance}

  CH results from the dissociative recombination of CH$_3$$^+$ and is
mainly destroyed by photo-dissociation. CH$_3$$^+$ is the result of
two successive proton exchanges with H$_2$ starting from CH$^+$.

\begin{figure}
\centerline{
\begin{tabular}{c}
\includegraphics[width=7cm,angle=+90]{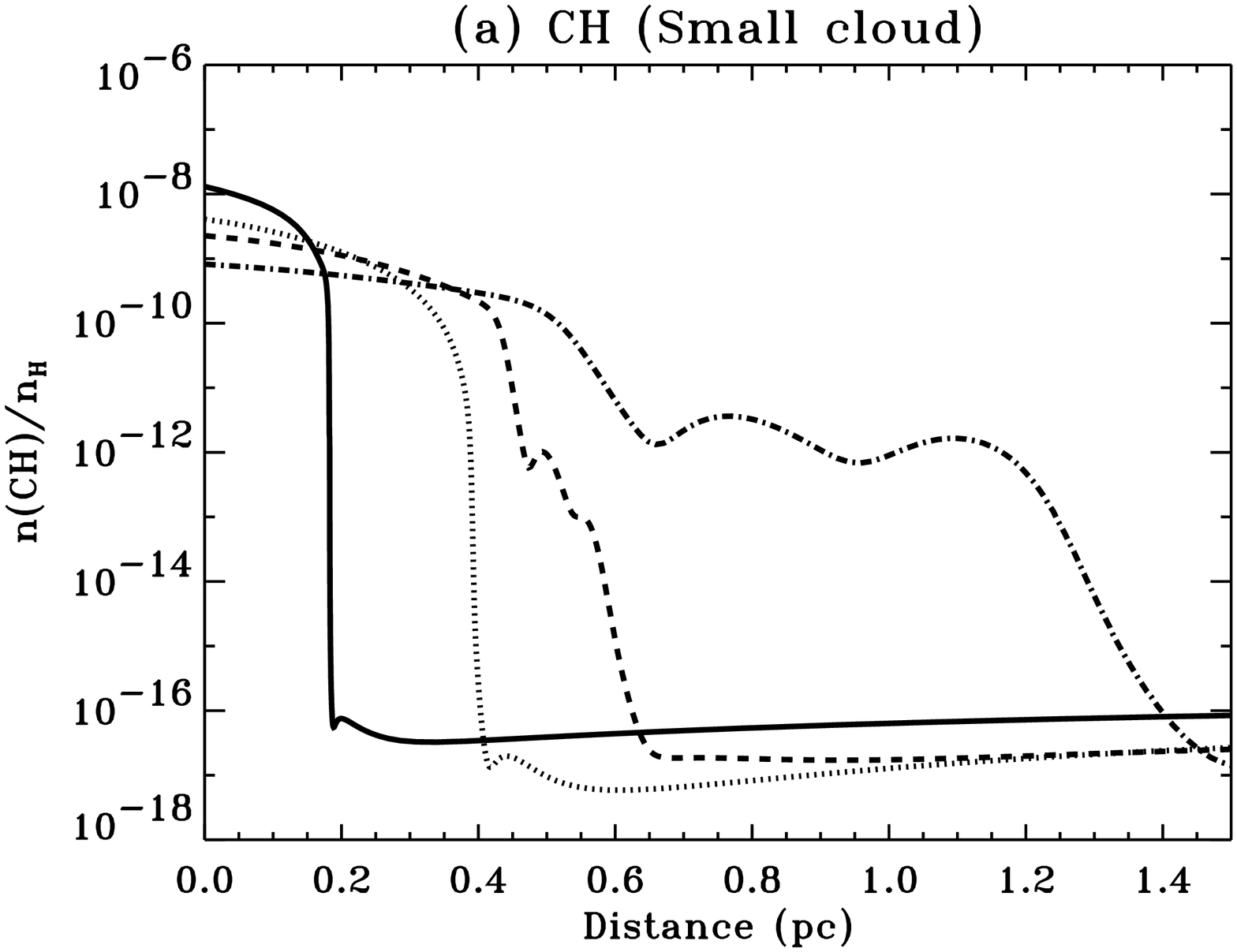}\\
\includegraphics[width=7cm,angle=+90]{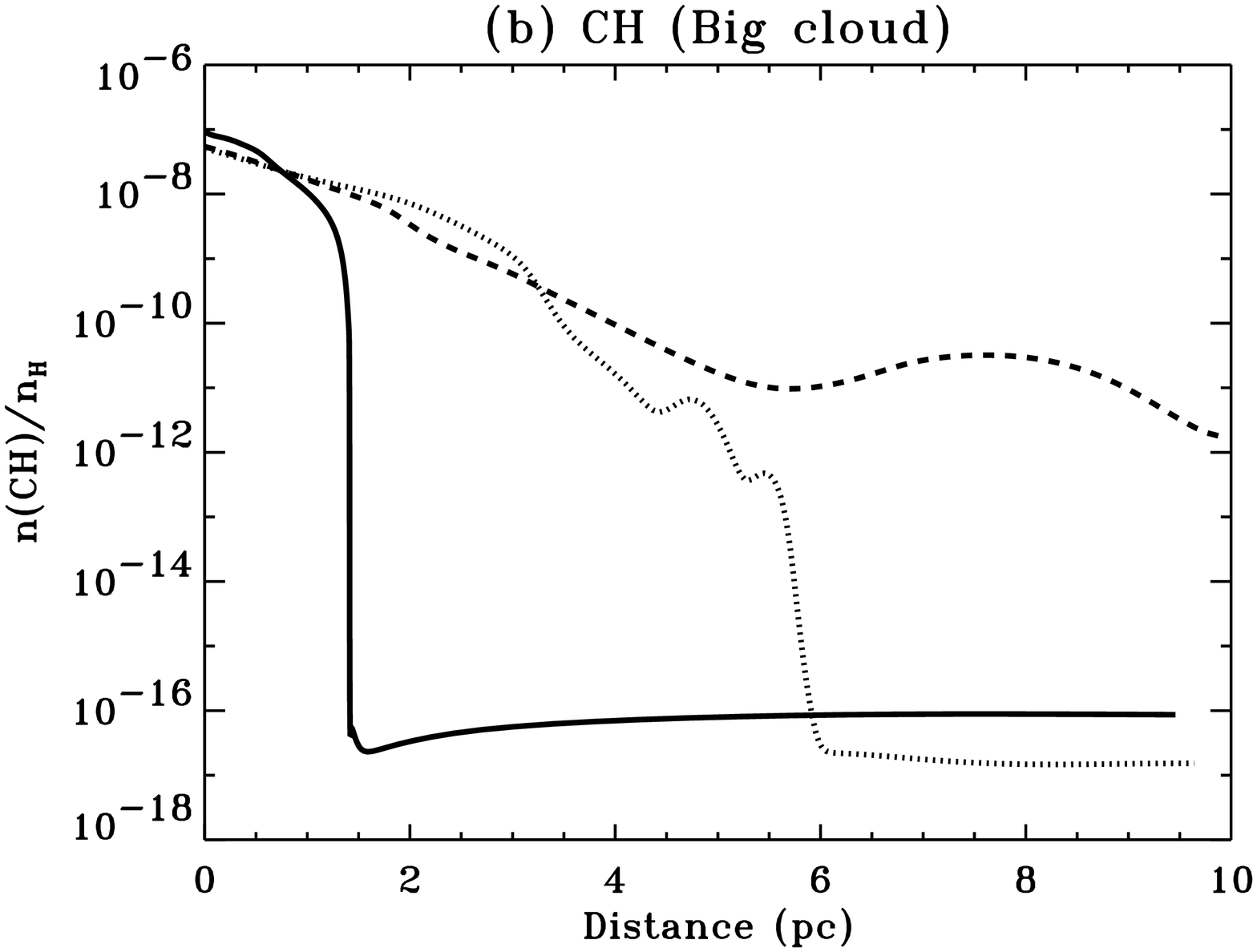}
\end{tabular}
} 
\caption{Abundance profile of CH for various diffusion
coefficients (same labels as figure \ref{Tn}).  }
\label{CH}
\end{figure}

  The decrease of CH in the cloud is related to the decrease of the H$_2$
 abundance (see figures \ref{H2} and \ref{CH}). In the diffusion
 front, the enhanced CH$^+$ abundance does not compensate for the higher
 temperature (which lowers the recombination rate) and the lower
 abundance of H$_2$.

  The activation temperature of 1.41~10$^4$~K for the reaction
C~+~H$_2~\rightarrow~$CH~+~H is too high to be overcome in the same
way as for CH$^+$. Hence there is a net decrease of CH column-density
when the diffusion is enhanced (the decrease comes from the lower
H$_2$ abundance in the cloud).

\subsubsection{Line profiles}

  We post-processed our simulations using a crude radiative transfer
  code in order to probe the effect of diffusion on the line profiles.
  We assume that a given species in each parcel of gas in the
  simulation emits (or absorbs) light with a Gaussian line shape in
  velocity space. The mean of the Gaussian is set to the local speed
  of the fluid, its width is set to the thermal speed of the species
  and its amplitude is set according to the column density of the
  parcel.  Note that we implicitly assumed that the turbulent velocity
  is a constant fraction of the local sound speed. In this simple
  picture such turbulent motions do not modify the shape of the
  spectra if the velocity scale is expanded appropriately. We hence
  neglect the turbulent Dopper shifts and concentrate only on the
  intrinsic line shapes which reflect thermal and chemical changes.

  In a first step, we don't take into account excitation effects.
Assuming an optically thin medium, or a line profile seen in
absorption, we compute the line shape across the box as the total
column-density of species $M$ per unit velocity bin centred on
velocity $v$:
\begin{equation}
\label{line}
N_v(M,v)=\int_{0}^{R}{\rm d}r~
N(M)\exp{[-~\frac{\mu(u-v)^2}{2 {\rm k} T}]}
\sqrt{\frac{\mu}{2 \pi {\rm k} T}}
\end{equation}
where $N(M)$ is the number density of species $M$, $\mu$ is its
molecular weight, $u$ is the local velocity of the fluid associated to
$M$ and $T$ its temperature. Note that the integral of $N_v(M,v)$ from
$v=-\infty$ to $v=+\infty$ yields the total column-density of species
$M$ across the simulation box. Taking turbulence into account could
be done simply by artificially decreasing $\mu$.

\begin{figure}
\centerline{
\begin{tabular}{c}
\includegraphics[width=7cm,angle=+90]{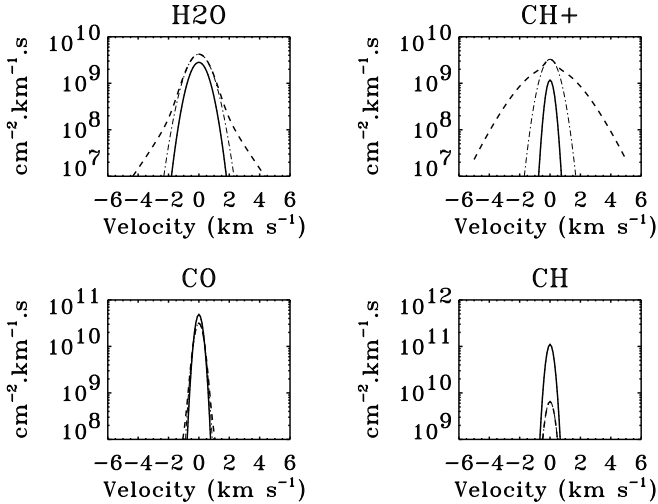}
\end{tabular}
} 
\caption{Line profile for H$_2$O, CH$^+$, CO and CH (see text).  The solid
line is in the case without turbulent diffusion, the dashed line is in
the case with turbulent diffusion ($L=8~\times~10^{-2}$ pc).  A
Gaussian line has been adjusted on the core of each profile (for
velocities lower then 0.2 km.s$^{-1}$). It is displayed as a dotted
line for the case without diffusion and as a dash-dotted line for the
case with diffusion.  In most cases, the dotted line cannot be
distinguished from the solid line.  }
\label{lines}
\end{figure}

\begin{figure}
\centerline{
\begin{tabular}{c}
\includegraphics[width=7cm,angle=+90]{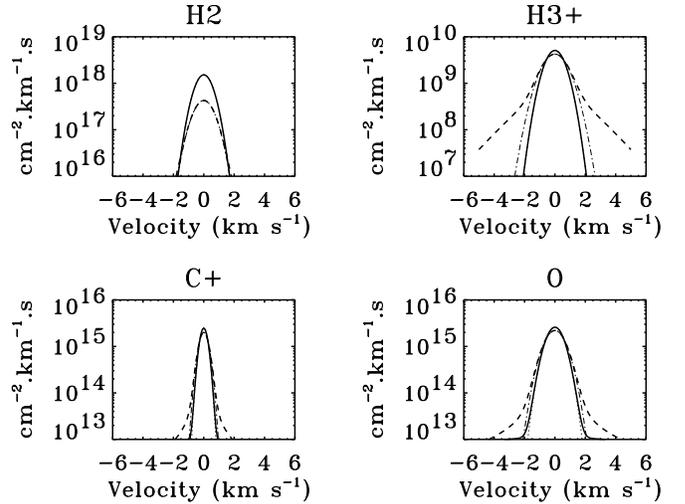}
\end{tabular}
} 
\caption{Line profile for H$_2$, H$_3$$^+$, C$^+$ and O (see text). Labels are the
same as in figure \ref{lines}.  }
\label{lines2}
\end{figure}

  Figures \ref{lines} and \ref{lines2} display $N_v$ for a few
molecules and atoms in the case without turbulent diffusion and in a
case with turbulent diffusion. Without turbulent diffusion, all
molecules show a narrow Gaussian profile corresponding to the temperature
of the CNM. Indeed, molecules are absent from the WNM. C$^+$ and O
on the contrary are equally abundant in both phases and they
show a wider secondary Gaussian feature, corresponding to emission
from the WNM. The amplitude of this second Gaussian is
smaller because there is much less total column-density in the WNM than
in the CNM.

  When turbulent diffusion is present, molecules are formed in the
transition layer at intermediate temperatures between the CNM and the
WNM.  This enhances the contribution of warm gas in the integral and
gives rise to broad line wings. The effect is strongest in the case of
CH$^+$ which has the most efficient yield due to turbulent
diffusion. The effect is inexistent for CH which subsists in the cold
phase only. This provides a potentially powerful comparative probe
between CH$^+$ and CH. Indeed, CH$^+$ lines are broader than CH and CO
lines in the diffuse ISM \citep{P04}.

\begin{figure}
\centerline{
\begin{tabular}{c}
\includegraphics[width=7cm,angle=+90]{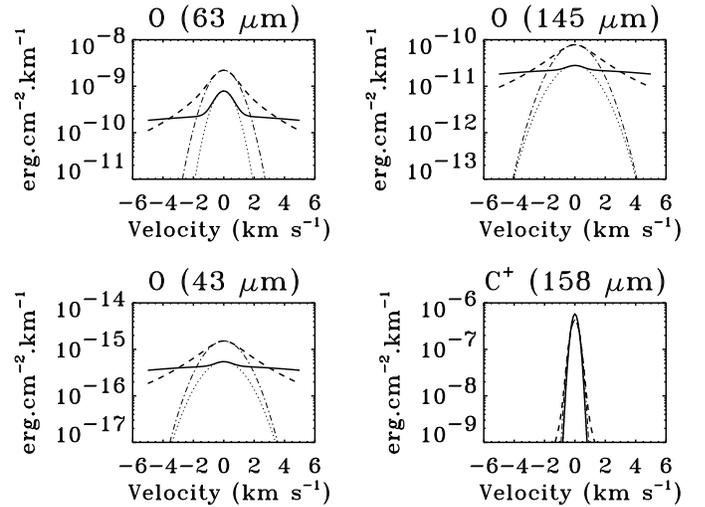}
\end{tabular}
} 
\caption{Emissivity profiles for the 63~$\mu$m (227~K), 145~$\mu$m
(98~K) and 43~$\mu$m (326~K) fine structure transitions of O and the
158~$\mu$m (92~K) fine-structure transition of C$^+$ (see text).
Labels are the same as in figure \ref{lines}.  }
\label{lines3}
\end{figure}

  We now compute the emission profiles of a few atomic transitions for
which our code provides the level populations at steady-state (they
are used to compute the atomic cooling functions). We integrate
equation (\ref{line}) using the local emissivities of these
transitions instead of the quantity $N(M)$. The results are displayed
in figure \ref{lines3}. The fine-structure transitions of O show a
strong effect with nearly exponential wings appearing in the case with
turbulent diffusion. O has a nearly constant abundance across the
simulation box. The wing effect in this case is entirely due to the
structure in density and temperature. The fine structure transitions
have energy temperatures of 227~K, 98~K and 326~K intermediate between
the CNM and WNM temperatures. Besides, the critical density for O is
much larger than the CNM density. It turns out that the emissivity of
these O lines is highest in the intermediate temperature and density
region inside the front. The change of integrated emissivity in the
line hence reflects only the change in total column-density found in
this intermediate region only. O line profiles are a good probe of the
thermal structure of such condensation fronts broadened by turbulent
transport of heat. On the other hand, the 158~$\mu$m (92~K) transition
of C$^+$ has a much lower critical density and its intensity is
highest in the CNM : its overall line profile barely sees the
structure of the front.

\section{Discussion}

\subsection{Initial conditions}
Our assumption of a closed box model determines the total mass
contained in our simulations. In particular the column-density of
hydrogen nuclei is determined by our initial conditions (ie: $8 \times
10^{19}$ cm$^{-2}$ in the standard runs and $8 \times
10^{20}$~cm$^{-2}$ for the big cloud runs). Then, the properties of the
radiation field at the outer side of the box completely determine the
final steady-state. Because our 1D simulations are unable to
self-consistently generate the turbulent motions, the value of the
diffusion coefficient also has to be treated as an external parameter.
 
  As has been detailed above, the total molecular enhancements due to
turbulent diffusion can either take place in the whole cloud or
be located at the interface.  The decrease of H$_2$ (and hence CH) is
caused by the lowered density of the cold cloud which happens over the
whole cloud. The enhanced production of CO, OH and H$_2$O due to
diffusion of H$^+$ ions from the WNM to the CNM occurs over the
whole cloud as well.  Finally H$_3^+$ and CH$^+$ production needs high
temperature {\it and } presence of H$_2$ molecules which can happen
only at the interface.  We therefore expect different scalings for
these effects with respect to the size of the cloud.

\label{bigcloud}
  In order to probe the effect of increasing the size of the cloud on
these global enhancements, we computed big cloud models with 10 times
as much mass as in our standard run. We simultaneously increased the
turbulent diffusion coefficient to account for the bigger size of the
cloud: bigger clouds should exist in regions where the correlation
length of the turbulence is larger. We investigated a factor 2
enhancement and a factor 10 enhancement of the diffusion coefficient
compared to the largest value we used in the small cloud case. In the
latter case, the simulation box of 10~pc was probably not large enough
to contain the wole cloud extended by turbulent diffusion and the WNM
phase almost disappears. In that last case, most of the volume
surrounding the cloud would be occupied by the gas at intermediate
temperature. Table \ref{table} summarises the results for some species
of interest. The molecular enhancements remain qualitatively the same
for the big cloud. However, the magnitude of the yields greater than 1
increases when the size of the cloud increases: non-linear effects in
the molecular enhancement make diffusion more efficient for the bigger
cloud.

  We also report in table \ref{table} the measured column densities
toward $\zeta$ Per. Although the models were not aimed at fitting the
$\zeta$ Per data, we obtain a relatively good agreement for most
species for the $C_{\rm t10}$ model, which is run with a standard CR
ionisation rate ($5\times10^{-17}$~s$^{-1}$). Diffusion considerably
enhances the size of the region where H$_3^+$ is present, with a
pathlength of 10~pc in this case. In particular, we produce reasonable
quantities of H$_3^+$ ($1.5~10^{13}$~cm$^{-2}$ for our best case)
without having to increase the cosmic ray ionisation rate as
observations sometimes seem to request \citep{MC02,LP04,SM06}. Our
models compared to the $\zeta$ Per observations also fall short of CO
and CH$^+$, both of which probably require the presence of an MHD
shock. Furthermore, in the case of the big cloud, CO photodissociation
may need a treatment for its self-shielding, not included in the present
simulations yet.

\begin{table*}
\begin{tabular}{lrrrrrrrrr}
\hline\hline
 Species     & $c$         & $c_{\rm t}$ & $y$       & $C$         & $C_{\rm t2}$& $y_2$      &$C_{\rm t10}$&$y_{10}$      & $\zeta$ Per\\
\hline
    H& 4.6(19)& 7.5(19)& 1.62& 1.1(20)& 1.8(20)& 1.66& 1.8(20)& 1.64& 5.7(20)\\ H$_2$& 2.0(19)& 6.9(18)& 0.34& 3.6(20)& 3.4(20)& 0.93& 3.5(20)& 0.96& 3.7(20)\\ C$^+$& 1.6(16)& 1.6(16)& 1.02& 1.4(17)& 1.5(17)& 1.05& 1.5(17)& 1.07& 1.8(17)\\
    C& 4.3(13)& 1.5(13)& 0.34& 8.1(15)& 4.1(15)& 0.51& 4.6(15)& 0.57& 2.9-3.6(15)\\
    O& 3.7(16)& 3.8(16)& 1.02& 3.6(17)& 3.6(17)& 1.02& 3.7(17)& 1.05& \ldots\\
   CO& 2.7(11)& 2.1(11)& 0.78& 2.3(13)& 9.5(12)& 0.41& 1.5(13)& 0.65& 5.4(14)\\
   OH& 4.8(11)& 2.1(12)& 4.40& 2.0(12)& 1.1(13)& 5.65& 1.7(13)& 8.53& 4.0(13)\\
H$_2$O/OH& 7.9(-2)& 3.6(-2)& \ldots& 8.0(-2)& 9.8(-2)& \ldots& 1.3(-1)& \ldots&\ldots\\
H$_3^+$& 7.5(10)& 9.5(10)& 1.27& 8.7(11)& 3.0(12)& 3.46& 1.5(13)&17.25& 8.0(13)\\
CH$^+$& 7.2(09)& 9.0(10)&12.56& 3.6(10)& 1.8(11)& 5.11& 6.2(11)&17.37& 3.5(12)\\
   CH& 6.1(11)& 4.1(10)& 0.07& 3.9(13)& 1.6(13)& 0.41& 1.8(13)& 0.46& 1.9-2.0(13)\\
H$_2$O& 3.8(10)& 7.6(10)& 2.00& 1.6(11)& 1.1(12)& 6.89& 2.1(12)&13.33& \ldots\\  
\hline
\end{tabular}
\caption{
Total column-densities (in cm$^{-2}$) of species of interest
with their relative yields. $c$ and $C$ are the column-densities
across the small and big cloud calculations without turbulent
diffusion.  $c_t$ is for the small cloud when maximum turbulent diffusion is
included with $L=8 \times 10^{-2}$~pc. $C_{\rm t2}$ and $C_{\rm
t10}$ are the results for the big cloud and turbulent diffusion with a
correlation length $L_2=2L$ and $L_{10}=10L$ respectively. The
relative yields are computed as $y=c_{\rm t}/c$, $y_2=C_{\rm t2}/C$
and $y_{10}=C_{\rm t10}/C$. Note that the yields are not strictly 1
for (H$+2$H$_2$), O and C$^+$ : this gives an estimate of the mass
conservation error in the code for individual
species. Column-densities as observed in the $\zeta$ Per line of sight
are provided in the last column. The last row diplays the ratio
of H$_2$O to OH column-densities. The last column contains the observed
column-densities as quoted by \cite{LP04}.
}

\label{table}
\end{table*}

\subsection{Observable effects on the lines}
   The effect on the {\it global} (ie: neglecting excitation effects)
 line profiles is not very strong, but the recent advances in
 observational techniques may allow the detection of subtle variations
 in the line shapes. However, for a typical molecule, rovibrational
 states provide a plethora of possible transitions that will be
 triggered at a variety of different temperatures.  

   We illustrated this effect for O: its line profile shows no diffusion
 effect when excitation is neglected (figure \ref{lines2}) but there
 is a clear difference for individual lines (figure \ref{lines3}).
 Such transitions should be good observational probes.  Furthermore, a
 slight increase in the wings of H$_2$ (see figure \ref{lines2}) could
 mean a huge impact for the lines found at intermediate temperatures
 in the front.  In fact, turbulent diffusion of H$_2$ towards warm gas
 could be an  explanation for the broadening of H$_2$ lines
 as seen by \cite{Lac05}.

 \subsection{Warm H$_2$}

  Several authors have reported observations of warm H$_2$ in cold
molecular gas \citep[see][ for example]{G02,F05}. Turbulent diffusion
could potentially be very efficient at increasing the column-density
of H$_2$ seen at intermediate temperatures, which would hence be seen
as warm H$_2$. An additional hint that turbulent diffusion might be
relevant in this problem is the presence of CH$^+$ in the line of
sights with warm H$_2$ \citep{FJ80,LD86}. Models of photon dominated
regions (PDR) with turbulent diffusion should hence account much
better for the observed H$_2$ excitation diagrams as published by
\cite{G02} and \cite{F05}.

   However, the computation of the whole spectrum needs to be carried
 out with a multigroup radiation transfer code in order to assess this
 effect quantitatively. This has been done already in the framework of
 steady-state PDRs computations which on the other hand lack a self
 consistent model for their hydrodynamical background. Including
 diffusion in these simulations could be one way of proceeding
 further. Note that diffusion would partially remove the uncertainty
 on the level population of H$_2$ newly formed on grains since in the
 case of turbulent diffusion, H$_2$ production is quite often
 dominated by transport processes.


\subsection{CH$^+$ and CH formation}

  Spectroscopic measurements of CH$^+$ and CH usually quote
column-densities of the order of 10$^{12}$ to 10$^{13}$~cm$^{-2}$ for
each velocity component, with CH and CH$^+$ approximately correlated
\citep{Lam90,P93,A94,CF04,P04,P05}. Such high amounts of CH$^+$ were a
puzzle for chemical modelers because the conditions for efficient
CH$^+$ formation were hard to be met in molecular gas. Since then,
several propositions were made to overcome this problem. Drift
velocities are an elegant way of increasing the reaction temperature
between charged and neutral species without necessarily increasing the
thermal energy of the gas: MHD shocks \citep{PF86} and MHD turbulence
dissipation in vortices \citep{J98} make use of this fact.

  The idea that CH$^+$ would be produced in turbulent boundary layers
at the interface between cold clouds and their surrounding medium was
originally suggested by \cite{D92}, based on steady-state chemistry
computations. Our simulations are the first ones to test this idea.
In our model we are unable to account for column-densities as high as
observed (the most we produce is 6.2~10$^{11}$~cm$^{-2}$ for our big
cloud run).  However, lines of sight with a small amount of CH$^+$ 
($<10^{12}$~cm$^{-2}$) and a normal amount of CH (2 10$^{13}$~cm$^{-2}$)
are also found \citep{G04,P04} which may be best explained by our
turbulent mixing scenario.

  CH$^+$ lines are broader than CH lines
\citep{Lam90,C94,Pr01,P05}. This might be a way to discriminate
between the various CH$^+$ formation channels. Indeed, MHD shocks
would produce little thermal broadening as the drift velocity is the
main factor used to overcome the activation temperature. Turbulence
dissipation will produce some line broadening as both heating and drift
motions in the vortices are needed to produce CH$^+$. Turbulent mixing
finally will produce the largest broadening as the gas has to be
readily brought up to temperature higher than the activation
temperature in order to produce CH$^+$.

  Finally, density probes such as the amount and excitation of C$_2$
\citep{G99,G04} produced may also be used to discriminate between the
various scenarii as the densities will tend to be high in MHD shocks
and low in CNM/WNM interfaces. However, we need to improve the present
work with an extensive chemical network in order to check whether
C$_2$ is indeed formed at the interface in association with CH$^+$ or
if it would reside in the cloud in our scenario. Note that our model
predicts that H$_3^+$ should correlate with CH$^+$ to some degree if
both are formed at the interfaces of CNM clouds. Unfortunately, the
present surveys of H$_3^+$ are too scarce to test this idea.

\subsection{Self-shielding}

As mentioned earlier, the Doppler parameter that controls the effect
of the broadening of the lines on the self-shielding of H$_2$
\citep{DB96} was implemented with a small value of 0.1 km.s$^{-1}$. We
ran another simulation without turbulence with a more reasonable value
of 1 km.s$^{-1}$ but the resulting total steady-state column-densities
were at most 7\% lower than in our reference run. This shows that such
a variation of the self-shielding function of H$_2$ has a small impact
on our results, as expected from the high level of saturation of H$_2$
lines.

\subsection{Diffusion coefficients}
We already discussed (see section \ref{turb}) the uncertainty on the
turbulent diffusion coefficients.  We concentrate here on the possible
difference between diffusion coefficients for different fluids and for
heat diffusion. As already mentioned, diffusion coefficients in a
multicomponent plasma are poorly known. Electrons should diffuse more
rapidly than the ions due to their much lower molecular weight, but
they are strongly coupled to the ions due to electrostatic forces. As
a result, it is likely that the effective diffusion coefficient for
the ions+electrons fluid is higher than the value $d_i$ we
adopted. Also, the long range of electrostatic forces helps
transferring energy without necessarily moving particles. As a
consequence heat diffusion coefficients are generally higher than
particle diffusion coefficients.

On the other hand, turbulent diffusion coefficients depend mainly on
the properties of the turbulent flow (as hinted at by section
\ref{turb}). As a result, turbulent heat diffusion and particle
diffusion are likely to proceed with the same coefficients. However,
MHD turbulence could provide different properties for the charged and
neutral fluids.  We ran simulations with enhanced diffusion in the
charged fluid that showed higher yields than in the case
$d_i=d_e$. Indeed, an enhanced diffusion coefficient in the charged
species introduces additional molecular production in the CNM via
H$^+$ transport from the hot to the cold medium because OH production
is highly sensitive to the H$^+$ ion density.  Magnetised shocks
\citep{PF86} and MHD turbulence \citep{J98} can also induce drift
velocities between the charges and the neutrals that may help the
production of CH$^+$-like species which involve reactions between the
two fluids and an activation temperature.

\subsection{Numerical diffusion}

  In any non-Lagrangian code, advection errors give rise to some
numerical diffusion.  The coefficient for numerical diffusion can be
estimated as a fraction of $D=v_r \Delta r$ where $v_r$ is the
relative velocity of the grid with respect to the gas and $\Delta r$
is the resolution of the grid.  In the present study, the resolution
is very high and grid velocities are very slow because the gas evolves
over chemical and diffusion time scales: the numerical diffusion is
always negligible except in the very early stages when the grid itself
finds its way to the diffusion front.

  Nevertheless we wish to warn hydrodynamical modelers
willing to include molecular reactions in simulations: they
should be careful with their numerical diffusion. Indeed, for strongly
dynamical flows, advection fluxes will be much greater than the
diffusion fluxes and spurious diffusion much greater than the
molecular diffusion might occur. In some cases, spurious molecular
production might result.


\section{Conclusions and future prospects}

  Turbulent diffusion affects the abundance of OH, H$_2$O, CH$^+$ and
H$_3^+$ the strongest.  In our most favourable case, the column-density
increases by more than a factor of 10 compared to the steady state
model. This shows that some dynamical effects may lead to significant
abundance variations, of comparable magnitude as uncertainties in
reaction rates, and should therefore not be ignored in ISM modelling.

  We have shown that H$_2$ mixed into hot gas can be the source of
warm molecular production thanks to reactions that otherwise do not
reach their activation temperature: we suggest this scenario as a new
possible channel for the formation of molecules such as CH$^+$ or
H$_3^+$ but also OH and H$_2$O in the big cloud case. Conversely, the
change of physical conditions within the CNM favour a higher H$^+$
abundance which can be the seed for OH and H$_2$O production (and to a
lesser extent, CO).  We verified that these effects are robust by
varying the parameters in our simulations such as the total
column-density of the cloud or the self-shielding function.

  Furthermore, line profiles of transitions with energy differences
intermediate between CNM and WNM temperatures can be considerably
affected even for species with no total column-density yield (such as
O). In particular, models for lines of sight containing warm H$_2$
should benefit from a turbulent diffusion treatment. In the future, a
combination of observational diagnostics should provide good
independent probes for the thermal, diffusion and chemical processes
at play in the transition regions between CNM and WNM.  For example,
some species like O or CH are insensitive to interface chemical
effects. O excitation is sensitive to thermal effects only.
Comparative spectroscopy between lines of O, CH, CH$^+$, H$_2$O could
hence be a very efficient tool to disentangle these different
phenomena. A typical signature of such diffusion fronts would be the
presence of CH$^+$ with little CH and wings in the line profiles of
CH$^+$, O and H$_2$O but not in CH nor C$^+$.
 
  Thanks to our multifluid treatment, we have also discovered a
marginal effect due to partial pressure equilibration. The higher
ionisation degree in the WNM is responsible for a large scale C$^+$
ion currents towards the cloud. These could in principle be the source
for some magnetic field and may also induce thermal inversions in the
temperature profiles of self-gravitating clouds.
  
  Finally, we want to insist on the necessity for simulations
including thermal and chemical effects to account for these turbulent
mixing effects. At a larger scale, 3D numerical simulations have
already been undertaken \citep[][for example]{E06} to probe turbulent
mixing effects on the interface between the warm ionised medium (WIM)
and the hot ionised medium (HIM). We wish to motivate multidimensional
computations of coupled hydrodynamics and chemistry that could help
calibrate the turbulent diffusion coefficients and assess the
conditions when a turbulent diffusion formalism is appropriate. In the
near future, such studies could design subgrid models for turbulent
mixing in astrophysical objects other than stars.

  In this work, we ran our simulations until they reach a steady
state, which may take up to 0.1~Gyr when turbulent diffusion is
included. This provides a technical point of reference which allows us
to filter all transient chemistry effects and leave only the turbulent
diffusion effects. However, the ISM is shaken by violent events such
as supernovae shocks on much shorter time scales of the order of 0.01 Gyr.
In a forthcoming paper, we plan to investigate the effect of
hydrodynamical and MHD shocks running through the structures studied
in this paper. We will hence be able to put a more consistent
dynamical picture behind the model of \cite{LP04}. With the additional
chemistry coming from compression and neutral/charge drift, we should
also improve our comparison to the observations.

\begin{acknowledgements}
  P.L. thanks G. Pineau des For\^ets and E. Falgarone for enjoyable as
well as enlightening discussions. P.L. acknowledges financial support
from the PCMI french national program and from the chaire d'excellence
at ENS. We also thank Pr. John Black for his detailed and
thorough but nevertheless positive and encouraging referee report
which led to significant improvement of the paper.

\end{acknowledgements}



\bibliographystyle{aa}
\bibliography{biblio}

\end{document}